\pdfoutput=1
\documentclass[preliminary,copyright,creativecommons]{eptcs}
\providecommand{\event}{EXPRESS/SOS 2026} 

\usepackage{iftex}

\ifpdf
  \usepackage{underscore}         \usepackage[T1]{fontenc}        \else
  \usepackage{breakurl}           \fi

\usepackage{bussproofs,
            mathtools,
            mathpartir,
            amsthm,
            amsmath,
            cleveref,
            tikz,
            xspace,
            eqparbox,
            float}
\usepackage[dvipsnames]{xcolor} 
\usepackage{annotate-equations}

\usepackage{etoolbox}
\newtoggle{extended}

\toggletrue{extended}

\newcommand{\appref}[1]{\iftoggle{extended}{Appendix~\ref{#1}}{the extended version~\cite{extended-version}}}

\tikzset{
  annotation/.append style={
    font=\rmfamily
  }
}

\DeclareMathAlphabet{\mathcal}{OMS}{cmsy}{m}{n}

\usetikzlibrary{positioning, arrows, arrows.meta}

\allowdisplaybreaks

\newcommand{\AXC}[1]{ \AxiomC{\shortstack{ #1 }} }
\newcommand{\UIC}[1]{ \UnaryInfC{\shortstack{ #1 }} }
\newcommand{\BIC}[1]{ \BinaryInfC{\shortstack{ #1 }} }
\newcommand{\TIC}[1]{ \TrinaryInfC{\shortstack{ #1 }} }

\usepackage{graphicx}
\usepackage{subcaption}

\usepackage{
    amsmath,
    amssymb,
    amsthm,
    stmaryrd,
    xcolor,
    xparse
}

\newcommand{\NSTs}{NSTs\xspace}
\newcommand{\pNSTs}{ASP\xspace}
\newcommand{\etal}{\emph{et~al.}\xspace}

\definecolor{nAct}{RGB}{168, 50, 92}
\definecolor{nTyp}{RGB}{145, 126, 42}
\definecolor{nPrc}{RGB}{95, 107, 122}
\definecolor{nLab}{RGB}{182, 50, 28}

\newcommand{\nAct}      [1] {\textcolor{nAct}{\mathtt{#1}}}
\newcommand{\nLbl}      [1] {{\textcolor{nAct}{\mathtt{#1}}}}
\newcommand{\nTyp}      [1] {\textcolor{nTyp}{\mathit{#1}}}
\newcommand{\nTypLbl}   [1] {\textcolor{nLab}{\mathtt{#1}}}
\newcommand{\nPrc}      [1] {\textcolor{nPrc}{\mathtt{#1}}}

\newcommand{\nT}            {\vdash}

\newcommand{\nC}[1][\Sigma] {\models_{#1}}

\newcommand{\nV}            {\mathcal{V}}
\newcommand{\nS}            {\mathcal{S}}

\newcommand{\nAlpha}        {\nTyp{\alpha}}

\newcommand{\nBeta}         {\nTyp{\beta}}

\newcommand{\nTheta}        {\nTyp{\theta}}

\newcommand{\nProc}     [2] {\nPrc{proc}(#1,#2)}
\newcommand{\nMsg}      [2] {\nPrc{msg}(#1,#2)}
\newcommand{\pred}{\longrightarrow}

\NewDocumentCommand{\nChoose}{mm}       {#1.\nAct{#2};~}
\NewDocumentCommand{\nCase}{mo}         {\nAct{case}~#1~(\IfValueT{#2}{#2)}}
\NewDocumentCommand{\nOpt}{m}           {\nAct{#1}\Rightarrow}
\NewDocumentCommand{\nEndCase}{}        {)}
\NewDocumentCommand{\nSend}{mm}         {\nAct{send}~#1~#2;}
\NewDocumentCommand{\nRecv}{mm}         {#1\leftarrow\nAct{recv}~#2;}
\NewDocumentCommand{\nClose}{m}         {\nAct{close}~#1}
\NewDocumentCommand{\nWait}{m}          {\nAct{wait}~#1;}
\NewDocumentCommand{\nFwd}{mm}         {#1 \leftrightarrow #2}
\NewDocumentCommand{\nPd}{mmO{}d<>}   {\IfNoValueTF{#4}
                                            {#1 \leftarrow #2[#3];}
                                            {#1 \leftarrow #2[#3]~#4;}}
\NewDocumentCommand{\nPdV}{mmO{}d<>}  {\IfNoValueTF{#4}
                                            {#1 \leftarrow #2[#3];}
                                            {#1 \leftarrow #2[#3]~\overline{#4};}}
\NewDocumentCommand{\nPdF}{mmO{}d<>}  {\IfNoValueTF{#4}
                                            {#1 \leftarrow #2[#3]}
                                            {#1 \leftarrow #2[#3]~#4}}
\NewDocumentCommand{\nPdVF}{mmO{}d<>}  {\IfNoValueTF{#4}
                                            {#1 \leftarrow #2[#3]}
                                            {#1 \leftarrow #2[#3]~\overline{#4}}}

\NewDocumentCommand{\nmLabel}{mmm}   {\nMsg{#1}{ \nChoose{#1}{#3} \nFwd{#1}{#2} }}
\NewDocumentCommand{\nmLabelR}{mmm}   {\nMsg{#1}{ \nChoose{#2}{#3} \nFwd{#1}{#2} }}
\NewDocumentCommand{\nmSend}{mmm}    {\nMsg{#1}{ \nSend{#1}{#3} \nFwd{#1}{#2} }}

\NewDocumentCommand{\nmClose}{m}     {\nMsg{#1}{ \nClose{#1} }}

\newcommand{\ntChoose} [1] {\oplus\{#1\}}
\newcommand{\ntOffer}  [1] {\&\{#1\}}
\newcommand{\ntOpt}    [1] {\nTypLbl{#1}\!:}
\newcommand{\ntSend}       {\otimes}
\newcommand{\ntRecv}       {\multimap}
\newcommand{\ntOne}        {\mathbf{1}}

 \usepackage{
    amsmath,
    amssymb,
    amsthm,
    stmaryrd,
    xcolor,
    xparse
}

\definecolor{mInstr}{RGB}{135, 95, 153}

\newcommand{\mInstr}      [1] {\textcolor{mInstr}{\mathtt{#1}}}

\newcommand{\mM}            {\mathcal{M}}

\newcommand{\mInc} [2][i] {#1: \mInstr{INC}(#2)}
\newcommand{\mDecj} [3][i] {#1: \mInstr{DECJ}(#2, #3)}
\newcommand{\mred}{\longrightarrow_{\mathtt{M}}}

 \usepackage{
    amsmath,
    amssymb,
    amsthm,
    stmaryrd,
    xcolor,
    xparse
}

\definecolor{menc}{RGB}{14, 54, 45}
\definecolor{tenc}{RGB}{99, 44, 25}

\newcommand{\mlb}   {\left\llbracket}
\newcommand{\mrb}   {\right\rrbracket}

\newcommand{\mli}   {\{\!\!|}
\newcommand{\mri}   {|\!\!\}}

\newcommand{\mlp}   {\llparenthesis}
\newcommand{\mrp}   {\rrparenthesis}

\newcommand{\etVal}[1]
    {\textcolor{tenc}{\mlp #1 \mrp}}

\newcommand{\etTree}[1]
    {\textcolor{tenc}{\mathcal{T}^\mM\mlp #1 \mrp}}

\NewDocumentCommand{\eReg}{mD<>{r}}
    {\textcolor{menc}{\mlb #1 \mrb_{#2}}}
\NewDocumentCommand{\eTree}{mD<>{t}}
    {\textcolor{menc}{\mli #1 \mri^\mM_{#2}}}
\NewDocumentCommand{\eInst}{mD<>{r_0,r_1}}
    {\textcolor{menc}{\mlb #1 \mrb^\mM_{#2}}}
\NewDocumentCommand{\eInstS}{mD<>{r_0,r_1,t}}
    {\textcolor{menc}{\mlb #1 \mrb^\mM_{#2}}}
\NewDocumentCommand{\eState}{m}
    {\textcolor{menc}{\mlb (#1) \mrb^\mM}}

\NewDocumentCommand{\eIncr}{mm}
    {\nmLabelR{#1}{#2}{incr}}
\NewDocumentCommand{\eDecr}{mm}
    {\nmLabelR{#1}{#2}{decr}}

\NewDocumentCommand{\eExp}{mmm}
    {\nmLabelR{#1}{#2}{exp_#3}}
\NewDocumentCommand{\eIns}{mmm}
    {\nmLabelR{#1}{#2}{ins_#3}}
\newcommand{\instsep}{\,|\,}

\newcommand{\mCM}   {\nC[\Sigma_\mM]}
\newcommand{\mSM}   {\Sigma_\mM}
 
\newtheorem{theorem}{Theorem}

\theoremstyle{definition}
\newtheorem{definition}{Definition}

\title{On the Expressiveness of \\Asynchronous Processes with Nested Session Types}
\author{Sjors Wortelboer \quad Dan Frumin \quad Jorge A. Pérez
\institute{University of Groningen, The Netherlands}
}
\def\titlerunning{On the Expressiveness of Asynchronous Processes with Nested Session Types}
\def\authorrunning{S. Wortelboer, D. Frumin \& J.A. Pérez}
\begin{document}
\maketitle

\begin{abstract}
Session types are a type-based approach to correct message-passing programs, where types specify protocols for channels and type checking guarantees communication correctness. 
Traditional session types express regular protocols only, which leaves out many protocols found in practice. 
Aiming at overcoming this limitation, Nested Session Types (\NSTs) can specify nested recursive protocols.  
\NSTs are known to be characterised by one-counter machines and deterministic context-free languages. 
In this paper, we study the expressiveness of 
\pNSTs (Asynchronous Session Processes), 
the process model associated with \NSTs as introduced by Das \etal.
We prove that \pNSTs is Turing complete by giving a correct encoding of two-counter machines into typed processes. Our encoding uses a non-trivial \emph{state-aware} representation of counters and instructions that is neatly governed by \NSTs. 
\end{abstract}
\section{Introduction}

Ensuring the correctness of the message-passing programs that underpin large-scale distributed systems is a fascinating and pressing challenge. We are interested in compositional approaches to program verification where \emph{types} specify   protocols for channels and \emph{type checking} 
statically guarantees communication correctness for implementations, specified as \emph{processes} in some suitable formalism. 
In this approach, \emph{expressiveness} is a primary concern, in two ways: 
first, we aim for rich type systems that can express communication structures found in real-world scenarios; second, since types soundly enforce correctness by constraining admissible program behaviours, we would like to determine the expressiveness of the resulting typed processes---which correct programs can be expressed in a typed setting and which cannot.
In this paper, we study the expressiveness of a process language with Nested Session Types (\NSTs) by Das \etal~\cite{DBLP:conf/esop/DasDMP21,das2020nestedsessiontypes}. \NSTs are among the most expressive type systems for concurrency proposed to date.

Session types specify the order and type of messages to be exchanged on a channel~\cite{DBLP:conf/concur/Honda93,DBLP:conf/esop/HondaVK98}. Traditional session types can express \emph{regular} protocols, as a communication action and a continuation protocol can be composed. However, many practical protocols are not regular;  examples include protocols for 
the (de)serialisation of tree-structured data~\cite{DBLP:conf/icfp/ThiemannV16}
and for
(buffered) message relaying~\cite{rfc4976}. The quest for more expressive session types led to the \emph{context-free} session types by Thiemann and Vasconcelos~\cite{DBLP:conf/icfp/ThiemannV16}. 
In context-free session types, generalised protocol composition combined with polymorphic recursion enables the specification of context-free protocols. \NSTs achieve high expressivity using different means, namely \emph{nested type constructors}, which allow recursive type parameters to act like a stack.

A crucial step in laying the foundations of non-regular session types is studying what we may refer to as their \emph{intrinsic expressiveness}: the properties of these infinite structures in themselves, independent of a target programming model. 
Important notions for verification (type duality, type equality, subtyping) concern types in isolation. 
In this respect, the theory of context-free session types is well-developed~\cite{DBLP:journals/iandc/AlmeidaMTV22,DBLP:conf/tacas/AlmeidaMV20,DBLP:conf/concur/SilvaMV23}; for NSTs, Das \etal prove that type equality is decidable and give a sound algorithm. 
Gay \etal~\cite{gay2022differentshadesinfinitesession} offer a unified treatment of intrinsic expressiveness results for recursive session types. They complement the results in~\cite{DBLP:conf/esop/DasDMP21,das2020nestedsessiontypes} and obtain a strict hierarchy where \NSTs are precisely characterised by one-counter machines and deterministic context-free languages, with decidable type equivalence and duality.

In this paper, we pursue a different yet complementary goal: establishing the expressiveness of the language underlying \NSTs, as introduced by Das \etal~\cite{DBLP:conf/esop/DasDMP21,das2020nestedsessiontypes}. We refer to this language as \pNSTs (Asynchronous Session Processes).
The language implements asynchronous communication in configurations, which contain processes; its formulation follows the Curry-Howard correspondence between intuitionistic linear logic and session types~\cite{DBLP:conf/concur/CairesP10}. As a result, well-typed configurations in \pNSTs enjoy confluence and deadlock freedom. 
However, the question of the expressiveness of well-typed \pNSTs configurations remains open.
Given the context-free nature of NSTs and the decidability of their key properties, we investigate whether the synchronisation mechanisms in \pNSTs, together with the intrinsic expressiveness of types, suffice to obtain Turing completeness.

Our main result answers this question affirmatively. We prove that \pNSTs is Turing complete by correctly encoding {Minsky machines}~\cite{minsky1967computation}. 
Minsky machines are a Turing-complete model that consists of two counters with a finite set of instructions; instructions can either increment a counter, decrement it (if the counter's current value is not zero), or jump to a given instruction (if the value is zero). 
With this minimal state and set of possible instructions, Minsky machines are, arguably, a simpler source for encoding, compared to Turing machines.
Let us briefly discuss how we model Minsky machines in \pNSTs.

Intuitively, counters can be implemented as stacks using polymorphic nested types to capture state at the type level. Given a natural  $m$, we define the nested type
    \begin{align*}
        \etVal{m} =&
        \begin{cases}
            0[]               & \text{if $m=0$} \\
            S[\etVal{m-1}]    & \text{if $m>0$}
        \end{cases}
    \end{align*}
    where $0$  and $S$  stand for  \emph{(parametric) type definitions} for zero and successor, given as follows:
    \[
    0[] \triangleq \ntOffer{
            \ntOpt{incr} S[0[]], 
            \ntOpt{jump} 0[] 
        }
        \qquad 
        S[\nAlpha] \triangleq~ \ntOffer{
            \ntOpt{incr} S[S[\nAlpha]],
            \ntOpt{decr} \nAlpha 
        }
    \]
    Expressed as a branching type ($\&$, external choice), this nested session type captures the protocol that specifies the correct state after executing an instruction, using nested types and polymorphic recursion. 
    This way, e.g., $0[]$ admits increment and jump (but not decrement) and $S[\nAlpha]$   admits increment and decrement (but not jump) by suitably modifying the nesting of the type variable $\nAlpha$, the parameter of~$S$.
Such counter types can then be combined on the process level (by a process that we will denote $\eReg{m}$) with the consequence that processes representing instructions are required to be \emph{state-aware}: they need a separate mechanism to keep track of the combined state of the two counters. This is where \NSTs enable a continuation of the state-aware processes, allowing transitions to be captured at the level of processes.

Overall, 
we believe that our encoding of Minsky machines is a significant test of the expressiveness of the coupling of \pNSTs and \NSTs. 
Indeed, we achieve a pleasant design in which \NSTs elegantly govern the correct behaviours of the encodings of counters and instructions, and their interactions.  

The rest of the paper is structured as follows. 
\Cref{s:back} recalls background notions of \pNSTs and \NSTs and the  definition of Minsky machines. 
The encoding of Minsky machines in \pNSTs is presented and illustrated in \Cref{s:encoding}; its correctness 
is given in terms of typability (\Cref{thm:typable}) and operational correspondence (\Cref{thm:soundness,thm:completeness}).
\Cref{s:concl} concludes the paper and discusses some related work.

\section{Background}
\label{s:back}

\subsection{Nested Session Types}

We first present Nested Session Types (\NSTs), then their associated process language, which we call Asynchronous Session Processes (\pNSTs), and finally we state the correctness properties ensured by typing. 

\paragraph{Nested Session Types}

\NSTs extend the Curry-Howard correspondence between session types and intuitionistic linear logic~\cite{DBLP:conf/concur/CairesP10} with type variables and type definitions. There is a finite signature $\Sigma$ that collects these type definitions. 
The syntax and operational interpretation of types is given next.

\begin{definition}[Nested Session Types, \NSTs]
    The language of nested session types (\NSTs) is defined by:
        \begin{align*}
        A,B,C ::=&~ 
        A\ntSend B & \text{send channel of type $A$, continue as $B$} \\
        &|~ A\ntRecv B & \text{receive channel of type $A$, continue as $B$} \\
        &|~   \ntChoose{\ntOpt{l} A_l}_{l\in L} & \text{send label $k\in L$, continue as $A_k$} \\
        &|~ \ntOffer{\ntOpt{l} A_l}_{l\in L} & \text{receive label $k \in L$, continue as $A_k$} \\
        &|~ \ntOne & \text{close channel} \\
        &|~ \nAlpha & \text{type variable} \\
        &|~ V[\nTheta] & \text{defined type name (in $\Sigma$)}
    \end{align*}
\end{definition}
Type variables (denoted $\nAlpha,\nBeta,\ldots$) enable the parametrisation of types. \NSTs do not make use of type variable binders; instead, type definitions act as binders for a sequence of type variables by which they are parameterised.
Type names are defined according to a type definition, denoted $V[\overline{\nAlpha}]\triangleq A$, where $\overline{\nAlpha}$ is a list of distinct variables $A$ can refer to. 
These definitions are stored in the signature $\Sigma$ and allow nested type construction.
The type $V[\nTheta]$ denotes the instantiation of a type name $V$, which equirecursively unfolds into $A$ with type parameters substituted using $\nTheta=\overline{A}/\overline{\nAlpha}$.
Any type that is not a type name is structural (e.g. $V_1[\nTheta_1]\ntSend V_2[\nTheta_2]$). Any type that is structural and not a type variable is contractive in the type definitions. Types without free type variables are called closed types.

A key result in~\cite{das2020nestedsessiontypes} is that type equality is decidable and has a sound algorithm. 
Type equality follows the judgment 
$\psi;\nV \nT_\Sigma A\equiv A'$
where $\nV$ denotes the free variables in $A$ and $A'$, and $\psi$ is a collection of \emph{closures} of the form $V[\theta] \equiv U[\sigma]$ used to terminate cyclic equivalences.

\paragraph{The Process Language}
We now introduce the process language \pNSTs. This is exactly the language 
introduced by Das \etal~\cite{DBLP:conf/esop/DasDMP21,das2020nestedsessiontypes}.
It is based on the intuitionistic variant of the Curry-Howard correspondence, under an asynchronous communication discipline.

\begin{definition}[Processes and Messages]
    Given a set of channel/names ranged over by $x, y, z, \ldots$, the syntax of \emph{processes} and is defined by:
    \begin{alignat*}{5}
        P,Q ::=
& ~\nSend{x}{y} P 
            & \text{send channel $y$ over $x$}
        \quad
        &|~ \nRecv{y}{x} P 
            & \text{receive a channel over $x$}
        \\
&|~ \nChoose{x}{l_i} P 
            & \text{send label $\mathtt{l_i}$ over $x$}
        \quad
        &|~ \nCase{x}[\nOpt{l_i} Q_i]_{l_i\in L} 
            & \text{receive a label over $x$}
        \\
&|~ \nClose{x} 
            & \text{closed channel $x$}
        \quad
        &|~ \nWait{x} P
            & \text{wait for closing of $x$}
        \\
&|~ \nPd{x}{f}[\nTheta]<\overline{y}> P ~~~
            & \text{process name $f[]$ in $\Sigma$}
        \quad
&|~ \nFwd{x}{y}
            & \text{forward $x$ along $y$}
    \end{alignat*}
    The syntax of \emph{messages} is defined by:
    \begin{align*}
        M ::=&~ \nSend{x}{y} \nFwd{x}{x'}
            ~~|~~         \nChoose{x}{k} \nFwd{x}{x'}
            ~~|~~ \nClose{x}
    \end{align*}
\end{definition}

The syntax of processes follows the syntax of types closely. 
Notable is the addition of process definitions $f[\nTheta]$ as a dual of nested type construction on the type level. A definition is instantiated, providing channel $x$ to the continuation process $P$ and making use of channels $\overline{y}$. Process definitions, similar to type definitions, are defined on the global signature $\Sigma$.
Messages co-exist with processes as atomic processes to enable asynchronous message passing. 
They lift sending actions from their processes such that they can be consumed by their dual. 
Messages preserve linearity by renaming their channel continuation, which matches the continuation of the process producing them.

\begin{definition}[Typing for Processes \& Messages]
    \label{def:type-rules-pst}
    Given an environment of type variables $\nV$, 
    the typing judgment for processes and messages is denoted 
    $\nV;\Delta \nT_\Sigma P :: (x:A)$, 
    with selected typing rules in \Cref{fig:type-rule-process-def} (\appref{apx:type-rules-pst} gives a full account).
\end{definition}
\begin{figure}[!t]
    \begin{mathpar}
\AXC{$\nV;\Delta,y:A\nT P :: (x:B)$}
        \RightLabel{$\multimap R$}
        \UIC{$\nV;\Delta\nT \nRecv{y}{x}P :: (x:A\ntRecv B)$}
        \DisplayProof
        
\AXC{$\cdot;\nV\nT A\equiv A'$}
            \AXC{$\nV;\Delta,x:B\nT Q :: (z:C)$}
        \RightLabel{$\multimap L$}
        \BIC{$\nV;\Delta,x:A'\ntRecv B,y:A\nT \nSend{x}{y}Q :: (z:C)$}
        \DisplayProof
        
\AXC{$(\forall l\in L)$}
            \AXC{$\nV;\Delta\nT P_l :: (x:A_l)$}
        \RightLabel{$\& R$}
        \BIC{$\nV;\Delta\nT \nCase{x}[\nOpt{l} P_l]_{l\in L} :: (x:\ntOffer{\nOpt{l} A_l}_{l\in L})$}
        \DisplayProof
        
\AXC{$(k\in L)$}
            \AXC{$\nV;\Delta\nT Q :: (x:A_k)$}
        \RightLabel{$\& L$}
        \BIC{$\nV;\Delta,x:\ntOffer{\nOpt{l} A_l}_{l\in L}\nT \nChoose{x}{k} Q :: (z:C)$}
        \DisplayProof
        
\AXC{$\overline{y':B'}\nT f[\overline{\nAlpha}] =  P_f  :: (x':B) \in \Sigma$}
            \AXC{$\cdot;\nV\nT\Delta'\equiv \overline{y:B'[\nTheta]}$}
            \AXC{$\nV;\Delta,x:B[\nTheta]\nT  Q :: (z:C)$}
        \RightLabel{def}
        \TIC{$\nV;\Delta,\Delta'\nT \nPd{x}{f}[\nTheta]<\overline{y}> Q :: (z:C)$}
        \DisplayProof
    \end{mathpar}
    \caption{Selected typing rules for \pNSTs (\Cref{def:type-rules-pst}).}
    \label{fig:type-rule-process-def}
\end{figure}

 Following the intuitionistic (two-sided) presentation of linear logic, duality is defined implicitly, where a type is provided by the process over the intuitionistic channel and used by a client process over a channel in their environment. As a consequence, there are two typing rules per typing connective. 

We discuss the rules. Receiving ($\ntRecv$) is implemented by the providing process by adding the channel $y$ to the continuation. 
The dual sending action by the client process removes the channel $y$ to preserve linearity. Furthermore, it relies on type equality on the type of channel $y$, such that the receiving process can use a structurally different but equivalent type.
Offering a choice ($\ntOffer{\cdot}$) is implemented as branching on the providing process, where the client process selects the branch. 
Process definitions (def) require a well-typed definition from  $\Sigma$. All types use the substitution $\nTheta$ where, similar to sending, dependent channels are matched using type equality to allow for better compatibility when composing processes. Finally, the continuation process simply matches the substituted type provided by the definition.

The operational semantics of the language is defined in terms of \emph{configurations}, given next.

\begin{definition}[Configurations: Syntax and Typing]
    The syntax of configurations is defined by:
    \begin{align*}
        \nS ::= \nProc{x}{P} ~|~ \nMsg{x}{M}  ~|~  (\nS_1,\nS_2)^x
    \end{align*}
The typing rules are as follows:
    \begin{mathpar}
            \AXC{$\cdot;\Delta \nT P :: (x:A)$}
        \RightLabel{proc}
        \UIC{$\Delta \nC \nProc{x}{P} :: (x:A)$}
        \DisplayProof

            \AXC{$\cdot;\Delta \nT M :: (x:A)$}
        \RightLabel{msg}
        \UIC{$\Delta \nC \nMsg{x}{M} :: (x:A)$}
        \DisplayProof

            \AXC{$\Delta_1 \nC \nS_1 :: (x:A)$}
            \AXC{$x:A,\Delta_2 \nC \nS_2 :: (y:B)$}
        \RightLabel{comp}
        \BIC{$\Delta_1,\Delta_2 \nC (\nS_1,\nS_2)^x :: (y:B)$}
        \DisplayProof
    \end{mathpar}
\end{definition}

Configurations lift processes and messages with only closed types to the configuration level, as protocols need to be fully executable. The typing rule for composition binds a channel provided by some configuration $\nS_1$ to some configuration $\nS_2$ that uses it. This is the classical rule for composition derived from linear logic; it defines a partial ordering of the components of the configuration, and any configuration that has the same partial ordering is equivalent. We omit this presentation of composition as the bindings in the encoding are clearly defined.

We may now define reduction on the level of configurations.

\begin{definition}[Reductions]
    \label{def:reductions-pst}
    The reduction semantics for configurations is denoted $\pred$, with selected rules 
    in \Cref{fig:example-reductions-pst} (see \appref{apx:reductions-pst} for details).
\end{definition}
\begin{figure}
    \begin{align*}
        \nProc{y}{\nSend{x}{w} P}
            & \pred \nmSend{x}{x'}{w},~ \nProc{y}{P[x'/x]}
            & (\multimap S) \\[1.15ex]
        \nProc{x}{\nRecv{w'}{x} Q}, \qquad 
        &&
        \\
        \quad \nmSend{x}{x'}{w}
            & \pred \nProc{x'}{Q[x'/x,w/w']}
            & (\multimap C) \\[1.15ex]
        \nProc{d}{\nChoose{x}{k} P}
            & \pred\nmLabel{x}{x'}{k},~ \nProc{d}{P[x'/x]}
            & (\& S) \\[1.15ex]
        \nProc{x}{\nCase{x}[\nOpt{l} Q_l]}, \qquad 
        &&
        \\
        \quad \nmLabel{x}{x'}{k} 
            & \pred \nProc{x'}{Q_k[x'/x]} 
            & (\& C) \\[1.25ex]
        \nProc{z}{\nPd{x}{f}[\nTheta]<\overline{d}> Q}
            & \pred \nProc{a}{P_f[a/x,\overline{d}/\overline{y}, \nTheta]},~
            \nProc{z}{Q[a/x]}
            & (\text{defC})
    \end{align*}
    \caption{Selected reduction rules for \pNSTs (\Cref{def:reductions-pst}).}
    \label{fig:example-reductions-pst}
\end{figure}

The rules in \Cref{fig:example-reductions-pst} illustrate the asynchronous nature of communication in \pNSTs. 
Sending actions reduce independently by producing a message; a corresponding receiving action consumes the produced message separately. Consequently, a send action does not have to wait for the message to be consumed, and the receiving action does not have to be ready. Messages are produced to the right if sent over the intuitionistic channel and to the left otherwise, with renaming of channels to ensure the correct order of consumption. This preserves linearity and ensures type preservation.
Similarly, process definitions produce a newly spawned process to the left of the continuation. Whenever the continuation $Q$ is omitted, forwarding is implied as a continuation of the current process, skipping the renaming.

\paragraph{Properties of Typed Configurations}
We state here some properties of well-typed configurations. 
First, protocols are invariant under reduction.

\begin{theorem}[Type preservation]
    \label{thm:nst_type_pres}
    For a well-typed configuration in $\Delta \nC \nS :: (x:A)$:
    \begin{enumerate}
        \item If $\nS \equiv \nS'$, then $\Delta \nC \nS' :: (x:A)$;
        \item If $\nS \pred \nS'$, then $\Delta \nC \nS' :: (x:A)$.
    \end{enumerate}
\end{theorem}

Due to the acyclic nature of composition in \pNSTs, inherited from linear logic, progress is guaranteed for internal channels. 
A process ready to receive, or a message over an external channel, is called \emph{poised}; configurations consisting of only poised components are also called poised.

\begin{theorem}[Progress]
    \label{thm:nst_progr}
Suppose
    $\Delta \nC \nS :: (x:A)$. Either $\nS$ is poised, or $\nS \pred \nS'$.
\end{theorem}

Finally, confluence emerges due to the sequential independent definitions of reductions.
The language supports deterministic choices only, and so reductions persist until consumed---they can always be applied, creating different orders of application with the same resulting process.

\begin{theorem}[Confluence]
    \label{thm:conf}
    Suppose a configuration $\nS$ in \pNSTs.
    If $\nS \pred^* \nS_1$ and $\nS \pred^* \nS_2$, then there exists $\nS'$ such that $\nS_1 \pred^* \nS'$ and $\nS_2 \pred^* \nS'$.
\end{theorem}

\subsection{Minsky machines}

To show Turing-completeness for \pNSTs we need an appropriate formalism. We consider \emph{Minsky machines}, which have a simpler structure than Turing machines.
Minsky machines are Turing-complete, and so any system able to simulate them is by extension Turing-complete. 
In the following, we often use `machine' as shorthand for `Minsky machine'.

\begin{definition}[Minsky machines~\cite{minsky1967computation}]
    \label{def:2minsky}
    A Minsky machine $\mM$ is defined as a set of instructions $\{(1:I_1),....,(n:I_n)\}$, where
    instructions are defined as:
    \begin{align*}
        I ::=&~ \mInstr{INC}(j) & \text{increment on counter $j\in\{0,1\}$} \\
            &|~ \mInstr{DECJ}(j, k) & \text{decrement or jump on counter $j\in\{0,1\}, k \in \mathbb{Z}^+$}
    \end{align*}

     A machine includes a program counter $p$ indicating the label of the instruction being executed. The machine starts with both counters set to $0$ and the program counter $p$ set to the first instruction. The machine stops whenever the program counter is set to a non-existent instruction, i.e., $p > n$.

    The \emph{state} of a  machine $\mM$ is defined as the tuple $(i,m_0,m_1)^\mM$, where $i\in \mathbb{Z}^+$ and $m_0,m_1\in \mathbb{N}$. The tuple refers to the current instruction $i$ and the counters with values $m_0$ and $m_1$. The state is initialised to zero values on the counters and instruction $i$ as associated with the machine $\mM$.

    The semantics of the machine is defined as a relation on states, denoted $\mred$, given next:
    \begin{center}
        \begin{prooftree}
                \AXC{$\mInc{j}$}
                \AXC{$m'_j=m_j+1$}
                \AXC{$m'_{1-j}=m_{1-j}$}
            \RightLabel{Incr}
            \TIC{$(i,m_0,m_1)^\mM \mred (i+1,m'_0,m'_1)^\mM$}
        \end{prooftree}

        \begin{prooftree}
                \AXC{$\mDecj{j}{k}$}
                \AXC{$m_j>0$}
                \AXC{$m'_j=m_j-1$}
                \AXC{$m'_{1-j}=m_{1-j}$}
            \RightLabel{Decr}
            \QuaternaryInfC{$(i,m_0,m_1)^\mM \mred (i+1,m'_0,m'_1)^\mM$}
        \end{prooftree}

        \begin{prooftree}
                \AXC{$\mDecj{j}{k}$}
                \AXC{$m_j=0$}
            \RightLabel{Jump}
            \BIC{$(i,m_0,m_1)^\mM \mred (k,m_0,m_1)^\mM$}
        \end{prooftree}
    \end{center}
\end{definition}

The instructions can only manipulate the state. In case of a successful instruction $i$ (increment or decrement), the appropriate counter is updated, and the machine moves to the next instruction $i+1$. Decrementing is only possible when the counter value is not zero; otherwise, the value remains unchanged, and the machine jumps to instruction $k$.

\section{Encoding Minsky Machines in \pNSTs} \label{s:encoding}

\begin{figure}[t]
    \centering
\vspace{0.5cm} 
\begin{equation*}
    \eqnmarkbox[Purple]{cc}{\eState{i,m_0,m_1}} \triangleq 
            \eqnmarkbox[RoyalBlue]{r0}{\eReg{m_0}<r_0>},~~
            \eqnmarkbox[RoyalBlue]{r1}{\eReg{m_1}<r_1>},~~
            \eqnmarkbox[OliveGreen]{instp}{\eTree{m_0,m_1}},~~
            \eqnmarkbox[BrickRed]{instproc}{\eInstS{i \instsep m_0,m_1}}
\end{equation*}
\vspace{0.95cm}
\annotate[yshift=1.75em,xshift=0.15em]{above,left}{cc}{{\rmfamily State (\Cref{def:state-encoding})}}
\annotatetwo[yshift=1.75em,xshift=16.5em]{above,right}{r0}{r1}{{\rmfamily Counters (\Cref{def:counter-encoding})}}
\annotate[yshift=-0.8em,xshift=-9.5em]{below,left}{instp}{{\rmfamily Instantiation process (\Cref{def:instantiation-encoding})}}
\annotate[yshift=-2.5em,xshift=-16.75em]{below,left}{instproc}{{\rmfamily Instruction process (\Cref{def:instruction-encoding})}}
    \caption{Roadmap to the encoding of Minsky machine state in \pNSTs.}
    \label{fig:enc}
\end{figure}

In this section, we show how to encode the state of a   machine $\mM$ into \pNSTs. 
The encoding shows that \pNSTs is Turing complete, and illustrates how \NSTs can express stateful types that govern the behaviour of the encoding at the process level. The starting point for our proposed encoding is $\etVal{m}$, the interaction protocol for the representation of the natural number $m$,  discussed in the introduction.

Given a machine $\mM$ with state $(i, m_0, m_1)$, we define a configuration in \pNSTs denoted $\eState{i,m_0,m_1}$.
As \Cref{fig:enc} shows, we consider encodings of the counters, an instruction process that is \emph{state-aware} (i.e., it depends on the state of the counters), as well as an \emph{instantiation process}.
As we will see, the encoding of counters by itself is not sufficient to accurately determine the next instruction, in particular when performing decrements; to address this, the instantiation process codifies sufficient information about  $\mM$ to correctly enable the next instruction. 
The correctness of our encoding follows from
\begin{itemize}
    \item the \emph{typability} of $\eState{i,m_0,m_1}$  (\Cref{thm:typable}) and 
    \item \emph{operational correspondence} results, which ensure that (i)~every step possible from $(i, m_0, m_1)$ is correctly mimicked by $\eState{i,m_0,m_1}$ (soundness, \Cref{thm:soundness}) 
and (ii) all steps arising from $\eState{i,m_0,m_1}$ correspond to steps enabled from $(i, m_0, m_1)$ (completeness, \Cref{thm:completeness}).
\end{itemize}
 
\smallskip

We first describe the stateful representation at the level of types, namely  the one-counter types  that capture the state of the counters (\Cref{def:one-counter-protocol}), and  the \emph{instruction instantiation type}, which enables the continuation of instructions when decrementing (\Cref{def:instantiation-protocol}). 
These types provide the appropriate interfaces for typing the processesin \Cref{fig:enc}: counters $\eReg{m_i}<r_i>$ (with $i \in \{0,1\}$), the instantiation process $\eTree{m_0,m_1}$, and the encoding of the instruction $\eInstS{i \instsep m_0,m_1}$.
These interfaces, in turn, are crucial to enable the safe composition required to obtain the configuration $\eState{i,m_0,m_1}$ given by \Cref{def:state-encoding}.

Given a   machine $\mM$, we will define an associated signature $\mSM$. 
In giving the different ingredients of the encoding, we will be presenting type and process definitions that we assume are present in  $\mSM$. 
Types and processes for the encoding are presented in \Cref{ss:ptypes} and \Cref{ss:procs}, respectively. 
The correctness of the encoding is established in \Cref{ss:opcorr}, whereas an example illustrating the role of the different components of the encoding is given in \Cref{ss:examp}.

\subsection{Stateful Protocols}
\label{ss:ptypes}
As already discussed,  a key idea in our encoding is using Peano's representation of the naturals as a protocol that  relies on nesting of type parameters in \NSTs.
We will have two protocols, one per counter, and so the challenge lies in coordinating their behaviour according to the instructions of the machine.

\begin{definition}[Stateful counter protocol]
\label{def:one-counter-protocol}
    Given a natural number $m$, its associated counter type is:
    \begin{align}
        \etVal{m} =&
        \begin{cases}
            0[]               & \text{if $m=0$} \\
            S[\etVal{m-1}]    & \text{if $m>0$}
        \end{cases}
    \end{align}

    with type definitions in $\mSM$:
    \begin{align*}
        0[] =&~ \ntOffer{
            \ntOpt{incr} S[0[]], 
            \ntOpt{jump} 0[] 
        } \tag{zero}
        \\
        S[\nAlpha] =&~ \ntOffer{
            \ntOpt{incr} S[S[\nAlpha]],
            \ntOpt{decr} \nAlpha 
        } \tag{successor}
    \end{align*}
\end{definition}

This protocol is self-explanatory: the choices $\nTypLbl{incr}$, $\nTypLbl{jump}$, and  $\nTypLbl{decr}$ correspond to the three possible reduction steps for a machine  (increments, decrements, jumps), while accounting for successive steps. This is preferable to having two labels corresponding to instructions (with sub-cases for decrement-and-jump). The nesting in types is evident, as is the fact that state does not change in the case of a jump.
As a convention, we use type variables $\nAlpha$ and $\nBeta$ as parameters for the first and second counter, respectively.

In principle, given an instruction $i$ of $\mM$, we would like to have an encoding of instructions $\eInst{i}$ with judgement 
\[
\cdot;r_0:\etVal{m_0}, r_1:\etVal{m_1}, \cdots \nT \eInst{i} :: (x:\ntOne)
\]
where $r_0, r_1$ are names where the protocol counters are implemented.
Clearly, we do not know the exact type of instruction $i$, whether it operates on $r_0$ or $r_1$, or whether it may result in a jump. Our encoding of instructions must then be \emph{state-aware}.
As such we define a process $\eInst{i \instsep m_0,m_1}<r_0,r_1>$, which depends on either $0[]$ or $S[\nAlpha]$ on $r_0$ and $r_1$. 
An issue then arises: the protocol becomes $\nAlpha$ in case of a decrement, which is insufficient to capture the state of the counter.  Indeed, without extra information, the next instruction cannot be instantiated for the decrementing case, as it depends on $0[]$ or $S[\cdot]$.

To overcome this difficulty, the {instantiation process} will act as a coordinator between the process representation of counters and instructions. 
We define a corresponding \emph{instantiation protocol} for instructions, able to match $\nAlpha$ after decrementing a counter as well as to trigger the next instruction, enabling interaction with both counters. 
To define the instantiation protocol, we first characterise all the instructions that may occur in the possible states of $\mM$; this is needed to define $\mSM$ appropriately.
We use the following auxiliary notion. 

\begin{definition}
    \label{def:reach}
   Given a machine $\mM$, the set of \emph{accessible instructions} $L_\mM$ is defined as 
    \[
        L_\mM = 
        \{i+1~|~(i:I_i)\in\mM \}
        \cup
        \{k~|~(\mDecj{j}{k})\in\mM\}
    \]
\end{definition}
The set $L_\mM$ contains indices for the instructions of $\mM$ and possible indices that lie out of bounds of $\mM$. It is not the set of all the reachable instructions, but an over-approximation of it.
It is easy to verify that the set $L_\mM$ is finite, as there are only finite instructions in $\mM$.

\begin{definition}[Instantiation protocol]
    \label{def:instantiation-protocol}
    Given $m_0,m_1\in\mathbb{N}$ and a machine $\mM$, the \emph{instantiation protocol} $\etTree{ m_0,m_1 }$ is given as:
    \begin{align}
        \etTree{ m_0,m_1 } = \begin{cases}
            T^{0,0}[]
            & \text{if $m_0=m_1=0$} \\
            T^{s,0}[\etVal{m_0-1}]
            & \text{if $m_0>0\wedge m_1=0$} \\
            T^{0,s}[\etVal{m_1-1}]
            & \text{if $m_0=0\wedge m_1>0$} \\
            T^{s,s}[\etVal{m_0-1},\etVal{m_1-1}]
            & \text{if $m_0>0\wedge m_1>0$}
        \end{cases}
    \end{align}
    with the type definitions in \Cref{fig:instantiation-protocol} included in $\mSM$.
\end{definition}
\begin{figure}[t]
    \begin{alignat*}{5}
        T^{0,0}[] =~& \ntOffer{~
            &&\ntOpt{exp_0} T^{s,0}[0[]],
            &&\ntOpt{exp_1} T^{0,s}[0[]], \\*
            &&&\ntOpt{dup} T^{0,0}[] \ntSend T^{0,0}[],
            &&\ntOpt{ter} \ntOne
        &&}
        \\*
        T^{s,0}[\nAlpha] =~& \ntOffer{~
            &&\ntOpt{ins_0} I[\nAlpha, 0[]], \\*
            &&&\ntOpt{exp_0} T^{s,0}[S[\nAlpha]],
            &&\ntOpt{exp_1} T^{s,s}[\nAlpha, 0[]], \\*
            &&&\ntOpt{dup} T^{s,0}[\nAlpha] \ntSend T^{s,0}[\nAlpha],
            &&\ntOpt{ter} \ntOne
        &&}
        \\*
        T^{0,s}[\nBeta] =~& \ntOffer{~
            &&\ntOpt{ins_1} I[0[], \nBeta], \\*
            &&&\ntOpt{exp_0} T^{s,s}[0[],\nBeta], \nBeta],
            &&\ntOpt{exp_1} T^{0,s}[S[\nBeta]], \\*
            &&&\ntOpt{dup} T^{0,s}[\nBeta] \ntSend T^{0,s}[\nBeta],
            &&\ntOpt{ter} \ntOne
        &&}
        \\*
        T^{s,s}[\nAlpha, \nBeta] =~& \ntOffer{~
            &&\ntOpt{ins_0} I[\nAlpha, S[\nBeta]],
            &&\ntOpt{ins_1} I[S[\nAlpha], \nBeta], \\*
            &&&\ntOpt{exp_0} T^{s,s}[S[\nAlpha], \nBeta],
            &&\ntOpt{exp_1} T^{s,s}[\nAlpha, S[\nBeta]], \\*
            &&&\ntOpt{ter} \ntOne
        &&}
        \\*
        I[\nAlpha, \nBeta] =~&
            \mathrlap{
            \nAlpha \ntRecv \nBeta \ntRecv \ntOne \ntRecv \ntOffer{\ntOpt{i} \ntOne}_{i\in L_\mM}
            }
    \end{alignat*}
    \caption{Type definitions for the  instantiation-protocol (\Cref{def:instantiation-protocol}).}
    \label{fig:instantiation-protocol}
\end{figure}
In this definition, the superscript in $T^{-,-}$ denotes the reflection of the zero/successor state of the one-counter types.
This agnostic protocol enables the handoff of the one-counter channels to some process abstracting the transition to jump at the process level. 
The choices that are associated with incrementing and decrementing are called \emph{expansion} ($\nTypLbl{exp}$) and \emph{instantiation} ($\nTypLbl{ins}$), respectively. 
Expansion predicts the next state, while instantiation simply receives the dependencies of the current instruction and the next instruction. 
The choices for duplication ($\nTypLbl{dup}$) and termination ($\nTypLbl{ter}$) are for bookkeeping purposes related to the decrementing tree data structure process that the protocol is implemented in.

\subsection{Processes}
\label{ss:procs}
Having presented the nested session types that govern our encoding, we turn our attention to processes.
We first give the encoding of the counter process, which closely follows the one-counter protocol; it  is a dependency of the instruction process.

\begin{definition}[Encoding a counter]
\label{def:counter-encoding}
    The encoding of a counter containing $m\in\mathbb{N}$ into a configuration $\nS$ in \pNSTs is given as follows, using  \Cref{def:one-counter-protocol}:
    \begin{align}
        \eReg{m} =&
        \begin{cases}
            \nProc{r}{ R^0 } & \text{if $m=0$} \\
            \eReg{m-1}<p>, \nProc{r}{ R^s[\etVal{m-1}/\nAlpha] } & \text{if $m>0$}
        \end{cases}
    \end{align}
    with process definitions in $\mSM$:
    \begin{align*}
        \cdot &\nT r^0[] = R^0 :: (r:0[]) \qquad \quad 
        p:\nAlpha \nT r^s[\nAlpha] = R^s :: (r:S[\nAlpha])
    \end{align*}
    and the constituent process definitions:
\begin{equation*}
    \begin{aligned}[t]
        R^0 = \nCase{r}
                &\nOpt{incr} \nPd{c}{r^0} \nPdF{r}{r^s}[0[]]<c>, \\*
                &\nOpt{jump} \nPdF{r}{r^0} 
            ~~\nEndCase
    \end{aligned}
    \qquad
    \begin{aligned}[t]
        R^s = \nCase{r}
                &\nOpt{incr} \nPd{c}{r^s}[\nAlpha]<p> \nPdF{r}{r^s}[S[\nAlpha]]<c>, \\*
                &\nOpt{decr} \nFwd{p}{r} 
            ~~\nEndCase
    \end{aligned}
\end{equation*}
\end{definition}

As mentioned before, while $\eReg{m_0}<r_0>$ uses type variable $\nAlpha$, the definition of $\eReg{m_1}<r_1>$ uses $\nBeta$.
Each counter is implemented as a stack, following the protocol definition. 
The continuation process for each choice reflects the construction of the continuation type. 
For increments, this is the re-instantiation of the current process followed by the successor. 
For jumps, we only re-instantiate the zero counter process. 
For decrements, the process providing $\nAlpha$ is forwarded.

The encoding of instructions is state-aware and meant to interact with the instantiation process:

\begin{definition}[Instruction processes]
\label{def:instruction-encoding}
    Given some instruction $i\in\mathbb{Z}^+$ of $\mM$, the encoding of $i$ into a configuration $\nS$ in \pNSTs depends on the values of the counters $m_0,m_1\in\mathbb{N}$; it is given as follows, using \Cref{def:one-counter-protocol,def:instantiation-protocol}:
    \begin{align*}
        \eInstS{i \instsep m_0,m_1} =&
        \begin{cases}
            \nProc{x}{ P^{0,0}_i } & \text{if $m_0=m_1=0$} \\
            \nProc{x}{ P^{s,0}_i[\etVal{m_0-1}/\nAlpha] } & \text{if $m_0>0 \wedge m_1=0$} \\
            \nProc{x}{ P^{0,s}_i[\etVal{m_1-1}/\nBeta] } & \text{if $m_0=0 \wedge m_1>0$} \\
            \nProc{x}{ P^{s,s}_i[\etVal{m_0-1}/\nAlpha,\etVal{m_1-1}/\nBeta] } & \text{if $m_0>0 \wedge m_1>0$}
        \end{cases}
    \end{align*}
    with the process definitions in $\mSM$ using \Cref{def:reach}:
    \begin{alignat*}{5}
        &r_0:0[],&&r_1:0[], &&t:T^{0,0}[], &&y:\ntOne 
            \nT p^{0,0}_{i\in L_\mM}[] &= P^{0,0}_i :: (x:\ntOne) \\
        &r_0:S[\nAlpha],~&&r_1:0[], &&t:T^{s,0}[\nAlpha], &&y:\ntOne
            \nT p^{s,0}_{i\in L_\mM}[\nAlpha] &= P^{s,0}_i :: (x:\ntOne) \\
        &r_0:0[], &&r_1:S[\nBeta],~&&t:T^{0,s}[\nBeta], &&y:\ntOne
            \nT p^{0,s}_{i\in L_\mM}[\nBeta] &= P^{0,s}_i :: (x:\ntOne) \\
        &r_0:S[\nAlpha], &&r_1:S[\nBeta], &&t:T^{s,s}[\nAlpha,\nBeta],~&&y:\ntOne
            \nT p^{s,s}_{i\in L_\mM}[\nAlpha,\nBeta] &= P^{s,s}_i :: (x:\ntOne)
    \end{alignat*}
    and the constituent process definitions are given in \Cref{fig:instruction-processes}.
\end{definition}
\begin{figure}[!t]
    \begin{align*}
        &P^{0,0}_{i} = \begin{cases}
            \eqmakebox[C1][l]{$
            \nChoose{r_0}{incr}
            \nChoose{t}{exp_0}
            \nPdF{x}{p^{s,0}_{i+1}}[0[]]<r_0~r_1~t~y>
            $}
                & \text{if $\mInc{0}$}
            \\
            \nChoose{r_1}{incr}
            \nChoose{t}{exp_1}
            \nPdF{x}{p^{0,s}_{i+1}}[0[]]<r_0~r_1~t~y>
                & \text{if $\mInc{1}$}
            \\
            \nChoose{r_j}{jump}
            \nPdF{x}{p^{0,0}_k}<r_0~r_1~t~y>
                & \text{if $\mDecj{j}{k}$}
            \\
            \nWait{y} Q^{0,0}_i
                & \text{if $i>n$}
        \end{cases} \\
&P^{s,0}_{i} = \begin{cases}
            \nChoose{r_0}{incr}
            \nChoose{t}{exp_0}
            \nPdF{x}{p^{s,0}_{i+1}}[S[\nAlpha]]<r_0~r_1~t~y>
                & \text{if $\mInc{0}$}
            \\
            \nChoose{r_1}{incr}
            \nChoose{t}{exp_1}
            \nPdF{x}{p^{s,s}_{i+1}}[\nAlpha,0[]]<r_0~r_1~t~y>
                & \text{if $\mInc{1}$}
            \\
            \eqmakebox[C1][l]{$
            \nChoose{r_0}{decr}
            \nChoose{t}{ins_0}
            \nSend{t}{r_0}
            \nSend{t}{r_1}
            \nSend{t}{y}
            \nChoose{t}{(i+1)} 
            \nFwd{t}{x}
            $}
                & \text{if $\mDecj{0}{k}$}
            \\
            \nChoose{r_1}{jump}
            \nPdF{x}{p^{s,0}_k}[\nAlpha]<r_0~r_1~t~y>
                & \text{if $\mDecj{1}{k}$}
            \\
            \nWait{y} Q^{s,0}_i
                & \text{if $i>n$}
        \end{cases} \\
&P^{0,s}_{i} = \begin{cases}
            \nChoose{r_0}{incr}
            \nChoose{t}{exp_0}
            \nPdF{x}{p^{s,s}_{i+1}}[0[],\nBeta]<r_0~r_1~t~y>
                & \text{if $\mInc{0}$}
            \\
            \nChoose{r_1}{incr}
            \nChoose{t}{exp_1}
            \nPdF{x}{p^{0,s}_{i+1}}[S[\nBeta]]<r_0~r_1~t~y>
                & \text{if $\mInc{1}$}
            \\
            \nChoose{r_0}{jump}
            \nPdF{x}{p^{0,s}_k}[\nBeta]<r_0~r_1~t~y>
                & \text{if $\mDecj{0}{k}$}
            \\
            \eqmakebox[C1][l]{$
            \nChoose{r_1}{decr}
            \nChoose{t}{ins_1}
            \nSend{t}{r_0}
            \nSend{t}{r_1}
            \nSend{t}{y}
            \nChoose{t}{(i+1)} 
            \nFwd{t}{x}
            $}
                & \text{if $\mDecj{1}{k}$}
            \\
            \nWait{y} Q^{0,s}_i
                & \text{if $i>n$}
        \end{cases} \\
&P^{s,s}_{i} = \begin{cases}
            \nChoose{r_0}{incr}
            \nChoose{t}{exp_0}
            \nPdF{x}{p^{s,s}_{i+1}}[S[\nAlpha],\nBeta]<r_0~r_1~t~y>
                & \text{if $\mInc{0}$}
            \\
            \nChoose{r_1}{incr}
            \nChoose{t}{exp_1}
            \nPdF{x}{p^{s,s}_{i+1}}[\nAlpha,S[\nBeta]]<r_0~r_1~t~y>
                & \text{if $\mInc{1}$}
            \\
            \eqmakebox[C1][l]{$
            \nChoose{r_j}{decr}
            \nChoose{t}{ins_j}
            \nSend{t}{r_0}
            \nSend{t}{r_1}
            \nSend{t}{y}
            \nChoose{t}{(i+1)} 
            \nFwd{t}{x}
            $}
                & \text{if $\mDecj{j}{k}$}
            \\
\nWait{y} Q^{s,s}_i
                & \text{if $i>n$}
        \end{cases} 
    \end{align*}
            \caption{Constituent processes used in the encoding of instructions (\Cref{def:instruction-encoding}).}
        \label{fig:instruction-processes}
    \end{figure}

In this definition, the subscripts $r_0$, ${r_1}$, and $t$ in the encoding refer to the internal renaming of the channels of the counter processes $0$ and $1$ and the instantiation process, in that order. 
Channels $x$ and $y$ never reduce; hence, no renaming is necessary. Note that $x$ is merely an artefact of the intuitionistic nature of the calculus and needs to be included.

The definitions for the instruction processes deal with branching by being state-aware, which is a consequence of the stateful protocols; an instruction must know when it can decrement or when it should jump, and deal with the subsequent state accordingly.
This highlights the need for the instantiation protocol/process: without it, the instruction cannot move forward.

In \Cref{fig:instruction-processes}, the instructions are implemented by sending the appropriate label associated with the current instruction action over the corresponding counter channel. 
For incrementing and jumping, it is possible to predict the next state; hence, we can instantiate the next instruction directly; for example, in the encoding of increments   as defined in $P^{0,0}_i$.
Note that for increments the state of the counters is synchronised with the instantiation process; this is not necessary for jumps, as the state remains unchanged.

The case for decrementing is slightly different after sending the label, as it is impossible to instantiate the next instruction over a generic type variable. 
Instead, the current instruction hands off the dependent processes and the next instruction index to the instantiation process, which closes the circle and allows the definitions to be well-typed; this occurs, for example, in the encoding of decrement-jump in $P^{s,s}_i$.

Lastly, there is a special case for out-of-bounds instructions. In this case, the machine cannot reduce further and has terminated. To simulate this, we use the external channel $y$, which halts further progress; for example, in the encoding of out-of-bound $i$ in $P^{0,0}_i$. The continuation process $Q_i$ can be any process as long as it is distinct for different $i$, to capture the final state for each possible value of $i$.

Similar to \Cref{def:counter-encoding}, process definitions are added to the global environment $\mSM$ to move to instructions and make use of $L_\mM$ (\Cref{def:reach}) to include the finite set of accessible instructions. Unlike the process definitions, which are defined on $L_\mM$, the encoding of an instruction is unbounded.

To continue to the next instruction after decrementing, we rely on the instantiation process, defined next. 
The definitions make use of a decrementing tree data structure, as it is the most straightforward both in implementation and in understanding the bookkeeping involved.

\begin{definition}[Instantiation process]
    \label{def:instantiation-encoding}
    Given some $m_0,m_1\in\mathbb{N}$,   the \emph{instantiation process}  $\eTree{m_0,m_1}<t>$ is a configuration $\nS$ in \pNSTs defined as follows, using \Cref{def:one-counter-protocol,def:instantiation-protocol,def:instruction-encoding}:
    \begin{align}
        \eTree{m_0,m_1}<z> =&
        \begin{cases}
            \nProc{z}{ N^{0,0} }                              
                & \text{if $m_0=m_1=0$} \\[0.1ex]
\eTree{0,0}<a>,~ \nProc{z}{ N^{1,0} }
                & \text{if $m_0=1 \wedge m_1=0$} \\[0.1ex]
\eTree{m-1,0}<a>,~ \nProc{z}{ N^{s,0}[\etVal{m_0-2}/\nAlpha] }
                & \text{if $m_0>1 \wedge m_1=0$} \\[0.1ex]
\eTree{0,0}<b>,~ \nProc{z}{ N^{0,1} }
                & \text{if $m_0=0 \wedge m_1=1$} \\[0.1ex]
\eTree{0,m_1-1}<b>,~ \nProc{z}{ N^{0,s}[\etVal{m_1-2}/\nBeta] }
                & \text{if $m_0=0 \wedge m_1>1$} \\[0.1ex]
\eTree{0,1}<a>,~ \eTree{1,0}<b>,~ \nProc{z}{ N^{1,1} }
                & \text{if $m_0=m_1=1$} \\[0.1ex]
\eTree{m_0-1,1}<a>,~ \eTree{m_0,0}<b>,~ \nProc{z}{ N^{s,1}[\etVal{m_0-2}]/\nAlpha] }
                & \text{if $m_0>1 \wedge m_1=1$} \\[0.1ex]
\eTree{0,m_1}<a>,~ \eTree{1,m_1-1}<b>,~ \nProc{z}{ N^{1,s}[\etVal{m_1-2}/\nBeta] }
                 & \text{if $m_0=1 \wedge m_1>1$} \\[0.1ex]
\eTree{m_0-1,m_1}<a>,~ \eTree{m_0,m_1-1}<b>, 
                & \text{if $m_0>1 \wedge m_1>1$} \\[0.1ex]
            \multicolumn{2}{@{}l}{\quad \nProc{z}{ N^{s,s}[\etVal{m_0-2}/\nAlpha, \etVal{m_1-2}/\nBeta] }}
        \end{cases}
    \end{align}
    with the process definitions in $\mSM$:
    \begin{alignat*}{4}
        \cdot&;\cdot&& &&\nT t^{0,0}[] &&= N^{0,0} :: (z:T^{0,0}[])
        \\
        \cdot&; a:T^{0,0}[]&& &&\nT t^{1,0}[] &&= N^{1,0} :: (z:T^{s,0}[0[]])
        \\
        \nAlpha&; a:T^{s,0}[\nAlpha]&& &&\nT t^{s,0}[\nAlpha] &&= N^{s,0} :: (z:T^{s,0}[S[\nAlpha]])
        \\
        \cdot&;&& b:T^{0,0}[] &&\nT t^{0,1}[] &&= N^{0,1} :: (z:T^{0,s}[0[]])
        \\
        \nBeta&;&& b:T^{0,s}[\nBeta] &&\nT t^{0,s}[\nBeta] &&= N^{0,s} :: (z:T^{0,s}[S[\nBeta]])
        \\
        \cdot&; a:T^{0,s}[0[]],&& b:T^{s,0}[0[]] &&\nT t^{1,1} &&= N^{1,1} :: (z:T^{s,s}[0[],0[]])
        \\
        \nAlpha&; a:T^{s,s}[\nAlpha,0[]],&& b:T^{s,0}[S[\nAlpha]] &&\nT t^{s,1}[\nAlpha] &&= N^{s,1} :: (z:T^{s,s}[S[\nAlpha],0[]])
        \\
        \nBeta&; a:T^{0,s}[\nBeta],&& b:T^{s,s}[0[],S[\nBeta]] &&\nT t^{1,s}[\nBeta] &&= N^{1,s} :: (z:T^{s,s}[0[],S[\nBeta]])
        \\
        \nAlpha,\nBeta&; a:T^{s,s}[\nAlpha,S[\nBeta]],~&& b:T^{s,s}[S[\nAlpha],\nBeta] &&\nT t^{s,s}[\nAlpha,\nBeta] &&= N^{s,s} :: (z:T^{s,s}[S[\nAlpha],S[\nBeta]])
    \end{alignat*}
\Cref{fig:examples-instantiation} gives selected cases of the constituent processes definitions; \Cref{fig:tree-node-dependencies} presents a dependency graph.
A full account is given in \appref{apx:instantiation-nodes}.
\end{definition}
\begin{figure}[!t]
    \centering
    \begin{tikzpicture}[
    ->,
    >=stealth,
    node distance=0.3cm and 1.4cm, 
    main/.style={circle, fill=OliveGreen!20, thick, minimum size=10mm},
]
\node[main] (t_00)                          {$N^{0,0}$};
    
    \node[main] (t_10) [below left=of t_00]     {$N^{1,0}$};
    \node[main] (t_01) [below right=of t_00]    {$N^{0,1}$};
    
    \node[main] (t_s0) [below left=of t_10]     {$N^{s,0}$};
    \node[main] (t_11) [below right=of t_10]    {$N^{1,1}$};
    \node[main] (t_0s) [below right=of t_01]    {$N^{0,s}$};
    
    \node[main] (t_s1) [below left=of t_11]     {$N^{s,1}$};
    \node[main] (t_1s) [below right=of t_11]    {$N^{1,s}$};
    
    \node[main] (t_ss) [below right=of t_s1]    {$N^{s,s}$};

\path (t_10) edge node [fill=white, inner sep=1pt] {$0$} (t_00);
    \path (t_01) edge node [fill=white, inner sep=1pt] {$1$} (t_00);

    \path (t_s0) edge node [fill=white, inner sep=1pt] {$0$} (t_10);
    \path (t_11) edge node [fill=white, inner sep=1pt] {$1$} (t_10);

    \path (t_11) edge node [fill=white, inner sep=1pt] {$0$} (t_01);
    \path (t_0s) edge node [fill=white, inner sep=1pt] {$1$} (t_01);

    \path (t_s0) edge [loop, out=245, in=205, looseness=6] node [below left] {$0$} (t_s0);
    \path (t_s1) edge node [fill=white, inner sep=1pt] {$1$} (t_s0);

    \path (t_s1) edge node [fill=white, inner sep=1pt] {$0$} (t_11);
    \path (t_1s) edge node [fill=white, inner sep=1pt] {$1$} (t_11);

    \path (t_1s) edge node [fill=white, inner sep=1pt] {$0$} (t_0s);
    \path (t_0s) edge [loop, out=295, in=335, looseness=6] node [below right] {$1$} (t_0s);

    \path (t_s1) edge [loop, out=245, in=205, looseness=6] node [below left] {$0$} (t_s1);
    \path (t_ss) edge node [fill=white, inner sep=1pt] {$1$} (t_s1);

    \path (t_ss) edge node [fill=white, inner sep=1pt] {$0$} (t_1s);
    \path (t_1s) edge [loop, out=295, in=335, looseness=6] node [below right] {$1$} (t_1s);

    \path (t_ss) edge [loop right] node {$0$,$1$} (t_ss);

\end{tikzpicture}     \caption{Instantiation process (\Cref{def:instantiation-encoding}): Dependencies between constituent process definitions indicating decrements over either the first ($0$) or the second $(1)$ counter.}
    \label{fig:tree-node-dependencies}
\end{figure}

{The instantiation process  $\eTree{m_0,m_1}<t>$ implements a tree-like structure where processes $N^{-,-}$ represent \emph{nodes}.}
The encoding builds naturally on the one-counter protocols, creating dependencies on states of decrements on the counters forming the tree structure, where the channels $a$ and $b$ denote whether the first or second counter is decremented, respectively.
The dependencies of the composition are presented in \Cref{fig:tree-node-dependencies}. 
Here, the superscript notation of \Cref{def:instantiation-protocol} is reused with the addition of $1$. In the context of the protocol, this annotation was used merely to denote state; in this definition, it is expanded to reflect the transitions between states.
\begin{itemize}
    \item $0$ refers to a zero state without transition;
    \item $1$ refers to the transition from a successor to a zero state;
    \item $s$ refers to the transition from a successor to a successor state.
\end{itemize}

Transitions are thus implemented as a dependency on a type definition, unlike the counter, which uses a generic type parameter for its dependency. Consequently, these processes are instantiated using the second-order predecessor, where counters and instructions depend on the first-order predecessor.

\begin{figure}[!t]
    \begin{align*}
        \begin{split}
N^{0,0} = \nCase{z}
                &\nOpt{exp_0} \nPd{c}{t^{0,0}} \nPdF{z}{t^{1,0}}<c> \\
                &\nOpt{exp_1} \nPd{c}{t^{0,0}} \nPdF{z}{t^{0,1}}<c> \\
                &\nOpt{dup} \nPd{c}{t^{0,0}} \nSend{z}{c} \nPdF{z}{t^{0,0}} \\
                &\nOpt{ter} \nClose{z}
            \nEndCase
        \end{split}
        \\
        \begin{split}
N^{1,0} = \nCase{z}
                &\nOpt{ins_0} \nRecv{r_0}{z} \nRecv{r_1}{z} \nRecv{y}{z} \\
                    &\qquad\nCase{z}
                        \nOpt{i} \nPdF{z}{p^{0,0}_i}<r_0~r_1~a~y>
                    \nEndCase_{i\in L\mM} \\
                &\nOpt{exp_0} \nChoose{a}{exp_0} \nPdF{z}{t^{s,0}}[0[]]<a> \\
                &\nOpt{exp_1} \nChoose{a}{dup}
                                    \nRecv{b}{a}
                                    \nChoose{a}{exp_1}
                                    \nChoose{b}{exp_0} 
                                    \nPdF{z}{t^{1,1}}<a~b>\\
                &\nOpt{dup} \nChoose{a}{dup} 
                                    \nRecv{a'}{a}
                                    \nPd{c}{t^{1,0}}<a>
                                    \nSend{z}{c}
                                    \nPdF{z}{t^{1,0}}<a'> \\
                &\nOpt{ter} \nChoose{a}{1} \nWait{a} \nClose{z}
            \nEndCase
        \end{split}
        \\
        \begin{split}
N^{s,s} = \nCase{z}
                &\nOpt{ins_0} \nChoose{b}{ter} \nWait{b}
                    \nRecv{r_0}{z} \nRecv{r_1}{z} \nRecv{y}{z} \\&\qquad\nCase{z}
                        \nOpt{i} \nPdF{z}{p^{s,s}_i}[\nAlpha,\nBeta]<r_0~r_1~a~y>
                    \nEndCase_{i\in L\mM} \\
                &\nOpt{ins_1} \nChoose{a}{ter} \nWait{a}
                    \nRecv{r_0}{z} \nRecv{r_1}{z} \nRecv{y}{z} \\&\qquad\nCase{z}
                        \nOpt{i} \nPdF{z}{p^{s,s}_i}[\nAlpha,\nBeta]<r_0~r_1~b~y>
                    \nEndCase_{i\in L\mM} \\
                &\nOpt{exp_0} \nChoose{a}{exp_0} \nChoose{b}{exp_0} \nPdF{z}{t^{s,s}}[S[\nAlpha],\nBeta]<a~b> \\
                &\nOpt{exp_1} \nChoose{a}{exp_1} \nChoose{b}{exp_1} \nPdF{z}{t^{s,s}}[\nAlpha,S[\nBeta]]<a~b> \\
                &\nOpt{ter} \nChoose{a}{ter} \nChoose{b}{ter} \nWait{a} \nWait{b} \nClose{z}
            \nEndCase
        \end{split}
    \end{align*}
    \caption{Examples of leaf, linear and binary nodes of the instantiation process (\Cref{def:instantiation-encoding}).}
    \label{fig:examples-instantiation}
\end{figure}

In \Cref{fig:examples-instantiation} we present the most representative constituent process nodes of the instantiation process, with the remaining nodes following similar constructions. We can differentiate nodes by the number of decrementing dependencies in the tree: leaf nodes where both counters are zero, linear nodes with one non-zero counter and binary nodes with both counters non-zero. We comment on some of them:

\begin{itemize}
    \item Node $N^{0,0}$ is a leaf that just captures the zero state for both counters. This node exists primarily due to the infinite recursive nature of the one-counters and forms the starting point from which the decrementing tree is constructed, both in the encoding and in the simulation.

    \item Node $N^{1,0}$ is a linear node that captures the state of $1$ for the first counter and the transition to $0$. All states are constant, and therefore, it uses only closed types. The non-zero state implements instantiation to enable progress after decrement, which simply carries over the dependencies of the previous instruction, followed by the instantiation of the next instruction as designated by the current instruction. In case of an increment on the zero state, the node becomes binary by creating a new branch, creating a duplicate to create the choices. 
    {Because the node represents a zero state for a counter and consequently is linear,} a possible parent can represent a zero state for the same counter and be linear. Hence, duplication is needed so the parent can introduce new branches.

    \item Node $N^{s,s}$ is a binary node that captures the state of incrementing state for both counters and the transition to a successor state for both. All states are variable expressions, and therefore, it uses open types; the process still needs to be instantiated with concrete values. 
    {For instantiation, it makes use of termination to remove the unused branch}, as residual processes need to be terminated in \pNSTs. Furthermore, because this is already a binary node, no new branches have to be created by either themselves or their parents; hence no implementation of duplication.
\end{itemize}

To manage the tree, process definitions are added to the global environment $\mSM$. Note that \Cref{def:instantiation-encoding} depends on the process definitions given in \Cref{def:instruction-encoding}.

\begin{figure}[!t]
    \centering
    \begin{tikzpicture}[
every node/.style={rectangle, fill, minimum size=8mm, font=\bfseries, inner sep=1pt},
edge from parent/.style={draw, -Latex},
level 1/.style={sibling distance=2.5cm},
    level 2/.style={sibling distance=1.5cm},
    level distance=1.3cm,
originalA/.style={fill=OliveGreen!20,edge from parent/.style={draw=blue!80!black }},
    originalB/.style={fill=RubineRed!20, edge from parent/.style={draw=blue!80!black }},
    originalC/.style={fill=Emerald!20, edge from parent/.style={draw=blue!80!black }},
transformed/.style={fill=black!15}
]

\node[originalA] (rootA) {$\eTree{2,0}<t>$}
        child { node[originalB] {$\eTree{1,0}<a_0>$}
            child { node[originalC] {$\eTree{0,0}<a_1>$} }
        };

\draw[-{Latex[width=3mm,length=3mm]}, line width=2pt, gray] 
        (1.5, -1.5) -- (3.5, -1.5) node[midway, above, draw=none, fill=none] {$\mathtt{exp}_1$};

\begin{scope}[xshift=7cm, yshift=0.75cm]
        \node[originalA] (rootD) {$\eTree{2,1}<t'>$}
child { 
                node[originalB] {$\eTree{1,1}<a_0>$}
                child {
                    node[originalC] {$\eTree{0,1}<a^0_1>$}
                    child { node[transformed] {$\eTree{0,0}<a^0_2>$} }
                }
                child {
                    node[originalC] {$\eTree{1,0}<b_1>$}
                    child { node[transformed] {$\eTree{0,0}<a^2_2>$} }
                }
            }
child {
                node[originalB] {$\eTree{2,0}<b_0>$}
                child {
                    node[originalC] {$\eTree{1,0}<a^1_1>$}
                    child { node[transformed] {$\eTree{0,0}<a^1_2>$} }
                }
            };
    \end{scope}

\end{tikzpicture}     \caption{Example of branching of a linear node in \Cref{def:instantiation-encoding}.}
    \label{fig:branching-example}
\end{figure}

\Cref{fig:branching-example} illustrates how the decrement tree branches. 
The root node starts by duplicating its branch and then updating each branch using expansion to represent the different ways of decrementing. In turn, the branch with non-zero states will do the same, while the branch with a zero state will remain linear and forward the expansion. The expansions are terminated in the leaf node that becomes linear and creates a new leaf node.

We are finally ready to encode the state of a machine, building upon the previous definitions.

\begin{definition}[Encoding state]
    \label{def:state-encoding}
Using \Cref{def:counter-encoding,def:instruction-encoding,def:instantiation-encoding}, the encoding of state for machine $\mM$ is defined as:
    \begin{align}
        \eState{i,m_0,m_1} = 
            \eReg{m_0}<r_0>,
            \eReg{m_1}<r_1>,
            \eTree{m_0,m_1},
            \eInstS{i \instsep m_0,m_1}
    \end{align}
\end{definition}

 \begin{figure}[t!]
    \centering
    \begin{tikzpicture}[
proc/.style={rectangle, minimum width=2cm, font=\bfseries\small},
lifeline/.style={gray!50, thick, dashed},
register/.style={rectangle, fill=RoyalBlue!20, inner sep=1.5pt, font=\small, minimum size=.75cm},
instruction/.style={rectangle, fill=BrickRed!20, inner sep=1.5pt, font=\small, minimum size=.75cm,},
instantiation/.style={rectangle, fill=OliveGreen!20, inner sep=1.5pt, font=\small, minimum size=.75cm},
action/.style={-Stealth, thick, RoyalBlue}
]
\node[proc] (Reg) {Counter Process $0$};
    \node[proc, right=1.5cm of Reg] (Ins) {Instruction Process};
    \node[proc, right=1.5cm of Ins] (Itr) {Instantiation Process};

\draw[lifeline] (Reg.south) -- +(0,-3.75) node(base1){};
    \draw[lifeline] (Ins.south) -- +(0,-3.75) node(base2){};
    \draw[lifeline] (Itr.south) -- +(0,-3.75) node(base3){};

\node[register] (R0) at (Reg |- 0,-1) {$\eReg{m_0}<r_0>$};
        \node[register] (R1) at (Reg |- 0,-2.25) {$\eReg{m_0+1}<r'_0>$};
        \draw[thick] (R0) -- (R1);

\node[instruction] (I0) at (Ins |- 0,-1) {$\eInstS{i \instsep m_0,m_1}<r_0,r_1,t>$};
        \node[instruction] (I1) at (Ins |- 0,-3.15) {$\eInstS{i+1 \instsep m_0+1,m_1}<r'_0,r_1,t'>$};
        \draw[thick] (I0) -- (I1);

\node[instantiation] (T0) at (Itr |- 0,-1.9) {$\eTree{m_0,m_1}<t>$};
        \node[instantiation] (T1) at (Itr |- 0,-3.15) {$\eTree{m_0+1,m_1}<t'>$};
        \draw[thick] (T0) -- (T1);

\draw[action] (I0) -- node[above, black, font=\scriptsize, sloped] {$\nLbl{incr}$} (R0);
    
\draw[action] (Ins |- T0) -- node[above, black, font=\scriptsize, sloped] {$\nLbl{exp_0}$} (T0);
    
\end{tikzpicture}     \caption{An increment on counter $0$.}
    \label{fig:increment-flow}
\end{figure}

\begin{figure}[t!]
    \centering
\begin{subfigure}{\textwidth}
    \centering
    \begin{tikzpicture}[
proc/.style={rectangle, minimum width=2cm, font=\bfseries\small},
lifeline/.style={gray!50, thick, dashed},
register/.style={rectangle, fill=RoyalBlue!20, inner sep=1.5pt, font=\small, minimum size=.75cm},
instruction/.style={rectangle, fill=BrickRed!20, inner sep=1.5pt, font=\small, minimum size=.75cm,},
instantiation/.style={rectangle, fill=OliveGreen!20, inner sep=1.5pt, font=\small, minimum size=.75cm},
action/.style={-Stealth, thick, RoyalBlue}
]
\node[proc] (Reg) {Counter Process $0$};
    \node[proc, right=1.5cm of Reg] (Ins) {Instruction Process};
    \node[proc, right=1.5cm of Ins] (Itr) {Instantiation Process};

\draw[lifeline] (Reg.south) -- +(0,-6.25) node(base1){};
    \draw[lifeline] (Ins.south) -- +(0,-6.25) node(base2){};
    \draw[lifeline] (Itr.south) -- +(0,-6.25) node(base3){};

\node[register] (R0) at (Reg |- 0,-1) {$\eReg{m_0}<r_0>$};
        \node[register] (R1) at (Reg |- 0,-2.25) {$\eReg{m_0-1}<r'_0>$};
        \draw[thick] (R0) -- (R1);

\node[instruction] (I0) at (Ins |- 0,-1) {$\eInstS{i \instsep m_0,m_1}<r_0,r_1,t>$};
        \node[instruction] (I1) at (Ins |- 0,-5) {$\eInstS{i+1  \instsep m_0-1,m_1}<r'_0,r_1,t'>$};
        \draw[thick] (I0) -- (I1);

\node[instantiation] (T0) at (Itr |- 0,-1.9) {$\eTree{m_0,m_1}<t>$};
\draw[thick] (T0) -- (Itr |- I1);

    \node[instantiation] (T2) at (Itr |- 0,-5.75) {$\eTree{m_0-1,m_1}<t'>$};

\draw[action] (I0) -- node[above, black, font=\scriptsize, sloped] {$\nLbl{decr}$} (R0);
    
\draw[action] (Ins |- T0) -- node[above, black, font=\scriptsize, sloped] {$\nLbl{ins_0}$} (T0);
    
\coordinate (r0Msg) at (Itr |- 0,-2.6);
        \draw[action] (Ins |- r0Msg) -- node[above, black, font=\scriptsize, sloped] {$r_0$} (r0Msg);
    
\coordinate (r1Msg) at (Itr |- 0,-3.1);
        \draw[action] (Ins |- r1Msg) -- node[above, black, font=\scriptsize, sloped] {$r_1$} (r1Msg);
    
\coordinate (yMsg) at (Itr |- 0,-3.6);
        \draw[action] (Ins |- yMsg) -- node[above, black, font=\scriptsize, sloped] {$y$} (yMsg);
    
\coordinate (ipMsg) at (Itr |- 0,-4.2);
        \draw[action] (Ins |- ipMsg) -- node[above, black, font=\scriptsize, sloped] {$\nLbl{(i+1)}$} (ipMsg);
    
\draw[action] (Itr |- I1) -- node[above, black, font=\scriptsize, sloped] {\textit{forward}} (I1);
    
\end{tikzpicture}     \caption{A decrement: Counter $0$ has value $m_0>0$.}
    \label{fig:decrement-flow}
\end{subfigure}

 \vspace{0.5em}

\begin{subfigure}{\textwidth}
    \centering
    \begin{tikzpicture}[
proc/.style={rectangle, minimum width=2cm, font=\bfseries\small},
lifeline/.style={gray!50, thick, dashed},
register/.style={rectangle, fill=RoyalBlue!20, inner sep=1.5pt, font=\small, minimum size=.75cm},
instruction/.style={rectangle, fill=BrickRed!20, inner sep=1.5pt, font=\small, minimum size=.75cm,},
instantiation/.style={rectangle, fill=OliveGreen!20, inner sep=1.5pt, font=\small, minimum size=.75cm},
action/.style={-Stealth, thick, RoyalBlue}
]
\node[proc] (Reg) {Counter Process $0$};
    \node[proc, right=1.5cm of Reg] (Ins) {Instruction Process};
    \node[proc, right=1.5cm of Ins] (Itr) {Instantiation Process};

\draw[lifeline] (Reg.south) -- +(0,-2.75) node(base1){};
    \draw[lifeline] (Ins.south) -- +(0,-2.75) node(base2){};
    \draw[lifeline] (Itr.south) -- +(0,-2.75) node(base3){};

\node[register] (R0) at (Reg |- 0,-1) {$\eReg{0}<r_0>$};
        \node[register] (R1) at (Reg |- 0,-2.25) {$\eReg{0}<r'_0>$};
        \draw[thick] (R0) -- (R1);

\node[instruction] (I0) at (Ins |- 0,-1) {$\eInstS{i \instsep 0,m_1}<r_0,r_1,t>$};
        \node[instruction] (I1) at (Ins |- 0,-2.25) {$\eInstS{k \instsep 0,m_1}<r'_0,r_1,t>$};
        \draw[thick] (I0) -- (I1);

\node[instantiation] (T0) at (Itr |- 0,-1) {$\eTree{0,m_1}<t>$};

\draw[action] (I0) -- node[above, black, font=\scriptsize, sloped] {$\nLbl{jump}$} (R0);
    
\end{tikzpicture}     \caption{A jump: Counter $0$ has value $m_0=0$.}
    \label{fig:jump-flow}
    \end{subfigure}
        \caption{Encoding of a decrement-and-jump  $\mDecj{0}{k}$ on counter $0$, as sequence diagrams.}
    \label{fig:inst}
\end{figure}
We use sequence diagrams to explain the encoding of the three possible steps of a machine.

\begin{enumerate}
    \item \Cref{fig:increment-flow} shows how an incrementing reduction is mimicked. 
    In the encoding, each process is able to predict the next state. 
    Consequently, the instruction process simply sends the instruction to the counter (as denoted by the arrow with label $\nTypLbl{incr}$) and then synchronises that change to the instantiation process (as denoted by the arrow with label $\nTypLbl{exp_0}$). Then, each process can move such that it matches the next state of the encoding. However, due to the concurrent, asynchronous nature of \pNSTs, some processes (in particular the instruction process) can move ahead. 
    Due to the properties of \pNSTs, like type preservation, lagging processes are able to catch up without side effects, which only means that the encoding simply skipped a step but is still consistent.

\item  \Cref{fig:decrement-flow} shows the representation of a decrement. 
Unlike incrementing, the instruction process is not able to predict the next concrete state; it relies on the instantiation process to move to the next instruction. 
This is highlighted by the sending of the dependencies, followed by the forward, where the root node takes the place of the current instruction. 
Furthermore, a branch is promoted to become the new root node. Unlike incrementing, the instruction process has to wait for the instantiation process to catch up before it can proceed to instantiate the next instruction.

\item Finally, \Cref{fig:jump-flow} shows how we mimic jumping. 
Similar to the incrementing instruction,  each process can predict the next state; in this case, however, the state does not change. 
Instead, the instruction is simply sent, and afterwards we move to the next instruction $k$.
\end{enumerate}

\subsection{Correctness}
\label{ss:opcorr}
Before illustrating our encoding (which we do in \Cref{ss:examp}), we address its correctness.
We first prove that the encoding is well-typed in \pNSTs; hence, it inherits type preservation, progress, and confluence.

\begin{theorem}[Typability]
    \label{thm:typable}
    Given any state $(i,m_0,m_1)^\mM$ we have: \begin{itemize}
        \item $\cdot \mCM \eReg{m_i} :: (r:\etVal{m_i})$ \;\; (for $i \in \{0,1\}$; cf. \Cref{def:one-counter-protocol,def:counter-encoding});
        \item $\cdot \mCM \eTree{m_0,m_1}<t> :: (t:\etTree{m_0,m_1})$ \;\; (cf. \Cref{def:instantiation-encoding,def:instantiation-protocol});
        \item $r_0:\etVal{m_0}, r_1:\etVal{m_1}, t:\etTree{m_0,m_1}, y:\ntOne \mCM \eInstS{i \instsep m_0,m_1} :: (x:\ntOne)$ \;\; (cf \Cref{def:instruction-encoding}).
    \end{itemize}
    Hence, by \Cref{def:state-encoding}:
    $$y:\ntOne \mCM \eState{i,m_0,m_1} :: (x:\ntOne).$$ 
\end{theorem}

Well-typedness of the encoding ensures not only that the properties of \pNSTs hold, but also that the encoding follows the protocols for the individual components, as correctness is partially captured on the level of types. 
\Cref{thm:typable} is proved by induction on the value of the counter and case analysis on the instruction; see \cite{sjors2026}
for the full typing derivations.

After establishing typability, the second step is to prove correspondence on the level of reductions.
To prove such operational correspondence, we show that the semantics machine is simulated correctly (soundness) and that the encoding does not introduce any side effects (completeness).
\begin{theorem}[Soundness]
    \label{thm:soundness}
    Let $(i,m_0,m_1)$ be a state for a 
    machine $\mM$. If $(i,m_0,m_1) \mred (i',m'_0,m'_1)$, then $\eState{i,m_0,m_1} \pred^+ \eState{i',m'_0,m'_1}$.
    \end{theorem}
The proof of \Cref{thm:soundness} follows largely from the construction of the stateful session types, which capture the semantics with regard to transitions of state for different instructions. As the encoding is well-typed (\Cref{thm:typable}) as long as the correct instruction is sent in correspondence with the current state $i$, the encoding behaves the same as the machine concerning soundness.

This leaves the validation of the correct instruction being executed. This is relatively easy to verify, as in the encoding, each instruction is prefixed with the sending of a correct label over the correct channel, with no further interaction between the instruction and counter process until the next instruction. The correctness of the next instruction follows in cases of incrementing and jumping, as the next instruction is immediately instantiated. The case of decrementing is different, as it makes use of the instantiation process; however, as such an instruction sends the label for the next instruction and the instantiation process takes this label to instantiate the correct next instruction, it is clear that, from this perspective, the encoding behaves correctly.

Because the encoding is well-typed, progress is guaranteed except over channel $y$, which is there to simulate termination for out-of-bound instructions.

 {The proof of soundness is further illustrated by the sequence diagrams in \Cref{fig:increment-flow,fig:inst}. Note: the diagrams focus on the sequence of the reductions that lead to the next state of an encoding such that it matches the one-step source reduction in the machine.}

\begin{theorem}[Completeness]
    \label{thm:completeness}
    Suppose $\eState{i,m_0,m_1} \pred^+ \nS$ where $\nS$ is a configuration in \pNSTs. Then there exists $(i',m'_0,m'_1)$ such that $\nS \pred^+ \eState{i',m'_0,m'_1}$ and $(i,m_0,m_1) \mred^+ (i',m'_0,m'_1)$.
\end{theorem}
The proof of \Cref{thm:completeness} follows the same structure as the proof of soundness (\Cref{thm:soundness}), using the fact that \pNSTs is confluent. Essentially, if a translation of the machine state reduces to some configuration $\nS$, then $\nS$ can further reduce to a translation of another machine state.
This follows from reordering of the reductions as a consequence of confluence and soundness.

As mentioned in the discussion of the sequence diagrams, it is entirely possible for a process like that of the instruction to move ahead without waiting for the counter to match the encoding first. In the context of \pNSTs each of these reductions can always be applied at a later state to match the encoding; at no point can other processes influence the reduction to the next state for such a process. Hence, again, this is simply a reordering of the order of reductions that was found in the proof of soundness.

\subsection{Examples}
\label{ss:examp}
To illustrate the behaviour of the encoding and our operational correspondence results, we consider the translation of a simple machine $\mM$:\[\{(\mInc[1]{1}),(\mInc[2]{0}),(\mDecj[3]{1}{2})\}.\]
Depending on the starting instruction, $\mM$ can exhibit both terminating and non-terminating 
behaviour:
\begin{itemize}
    \item Terminating computation: $(1,0,0) \mred (2,0,1) \mred (3,1,1) \mred (4,1,0) \not\mred$
    \item Non-terminating computation: $(2,0,0) \mred (3,1,0) \mred (2,1,0) \mred \cdots$
\end{itemize}
The first computation terminates because of the initial increment, which allows the decrement instruction to succeed and move out-of-bounds. In the second computation this initial increment is skipped; as a result, the decrement fails and jumps back to the initial instruction, allowing it to loop continuously.

{The sequence diagrams in \Cref{fig:increment-flow,fig:inst} help visualise how, for each step in the machine, a set of reductions in the encoding can be composed such that it leads to the encoded next state, justifying soundness for the encoding of the two computations. There is a difference: the terminating computation makes use of the instantiation process to move to the next instruction, while  the non-terminating computation does not use it, as the decrement fails and the state does not change.}

Because of the asynchronous nature of \pNSTs, the instruction process in the encoding can move independently from the other processes until it requires something of them in case of decrements. This can be illustrated by   the encoding of the terminating computation, which follows the flow of \Cref{fig:increment-flow}:
\begin{align*}
    \eState{1,0,0}\pred^+ ~
        &\eReg{0}<r_0>,~
            \eIncr{r'_0}{r_0}, \\*
        &\eReg{0}<r_1>,~
            \eIncr{r'_1}{r_1}, \\*
        &\underbrace{\eTree{0,0},~
            \eExp{t^1}{t}{1},~
            \eExp{t^2}{t^1}{0}}_{\nS_t},~  \eInstS{3 \instsep 1,1}<r'_0,r'_1,t^2>
\end{align*}
where $\pred^+$ denotes to the transitive closure of $\pred$.
Here, the encoding of the instruction is already at the state $(3,1,1)$. At this point, the consumption of all messages would allow the encoding to match for all other processes.
However, there is nothing preventing the process from not doing so, except when the instruction reduces further, following \Cref{fig:decrement-flow}:
\begin{align*}
    \eInstS{3 \instsep 1,1}<r'_0,r'_1,t^2> \pred^+ ~
        &\eDecr{r''_0}{r'_0},~ \underbrace{\eIns{t^3}{t^2}{1},~ \cdots,~ \nmLabelR{x}{t'}{4}}_{\nS_i}
\end{align*}

Because the protocol allows for a successful decrement, the current instruction hands over to the instruction process, as that process is aware of the state after decrementing the counter; this is done by sending the dependencies, the label for the next instruction and the channel that the instruction process offers.
The only reductions possible are those of the consumption of messages, in particular those of the instantiation process, as they construct the next instruction.
This continues the flow of \Cref{fig:decrement-flow}.
\begin{align*}
    \nS_t, \nS_i \pred^+ ~& \eTree{1,1}<t^2>,
        \eIns{t^3}{t^2}{1},~ \cdots ,~ \nmLabelR{x}{t'}{4} \\*
    =~& \eTree{0,1}<t_0>,~ \eTree{1,0}<t_1>,~ \nProc{t^2}{N^{1,1}},~
        \eIns{t^3}{t^2}{1},~ \cdots ,~ \nmLabelR{x}{t'}{4} \\*
    \pred^+~& \eTree{1,0}<t_1>,~ \eInstS{4 \instsep 1,0}<r''_0,r'_1,t_1>
\end{align*}

When we skip ahead to the instantiation steps (for branching, see \Cref{fig:branching-example}), then what is left is for the process associated to branch $0,1$ to terminate and instantiate the next instruction, using the branch $1,0$ and the dependencies received from previous instruction.
In this case, instruction $4$ is out-of-bounds: the instruction process becomes poised over channel $y$;  when the reductions of the counters take place (\Cref{fig:increment-flow,fig:decrement-flow}), {the encoding matches the machine state $(4,1,0)$}.

In the encoding of the non-terminating computation, this final synchronisation never happens; moreover, it never decrements, and so the instruction never makes use of the instantiation process. This means that increment and jump messages to the counter and the expand messages to the instantiation process just queue up. But at any point, they can be consumed such that the encoding can match the state of the simulated machine.

Notable in this example is that instructions never require anything from the counter and merely update their state. While at the process level the counters serve little function, their use comes from the session types they implement, enforcing the correct behaviour of processes using them. They essentially represent the possible reductions with state transitions independent of the definition of the instructions.

\section{Concluding Remarks}
\label{s:concl}
We have studied the expressiveness of \pNSTs, the asynchronous process language associated with the Nested Session Types by Das \etal. Starting from an intuitive encoding of numbers that naturally leverages nested types that implement Peano's representation of the naturals (\Cref{def:one-counter-protocol}), we encoded Minsky machines by developing instruction processes that are state-aware in order to precisely and correctly enact decrement-and-jump instructions. The correctness properties of our encoding are given in terms of typability (\Cref{thm:typable}) and operational correspondence (soundness and completeness guarantees, cf. \Cref{thm:soundness,thm:completeness}). 
Combined, these results entail that the coupling of \pNSTs and \NSTs is Turing complete.
In future work, we would like to explore connections between \pNSTs (and \NSTs) with other typed languages with logical foundations, such as CLASS~\cite{DBLP:conf/esop/RochaC23,DBLP:journals/corr/abs-2505-20848}, which features both polymorphism and state. 

\paragraph{Related Work} 
Closely related work has been discussed in the introduction; here we comment on other relevant literature.
Minsky machines are a convenient model for proving the Turing completeness of models of concurrency. They were used, for instance, by Busi \etal~\cite{DBLP:conf/icalp/BusiGZ03,DBLP:conf/icalp/BusiGZ04,DBLP:journals/mscs/BusiGZ09} to compare different forms of infinite behaviour in CCS and by Lanese \etal~\cite{DBLP:conf/lics/LanesePSS08,DBLP:journals/iandc/LanesePSS11} to establish the expressiveness of higher-order process calculi. In both cases, the calculi are untyped. 
In  {typed} process calculi, the situation is different, as types enforce correctness by considerably narrowing down the kind of (non-terminating) behaviours that well-typed processes can express. 
We are not aware of other typed encodings of Minsky machines.

In linearly-typed calculi, the limited flexibility is even more prominent, as linearity induces properties such as confluence and deadlock freedom. This is the case for typed frameworks derived from the Curry-Howard correspondence between session types and linear logic~\cite{DBLP:conf/concur/CairesP10}, which are strongly normalising. 
In this respect, the work of Balzer \etal~\cite{DBLP:conf/concur/BalzerPT18} is related: they consider a session-typed $\pi$-calculus with \emph{manifest sharing}, and show that adding recursive shared types to an otherwise linearly-typed language suffices to correctly encode the untyped, asynchronous $\pi$-calculus. 

Also in the typed setting, Kouzapas \etal~\cite{DBLP:conf/esop/KouzapasPY16,DBLP:journals/iandc/KouzapasPY19} obtain a series of relative expressiveness results concerning first- and higher-order $\pi$-calculi with session types (not based on linear logic). They discover that non-tail recursive types are essential to encode a name-passing calculus into a higher-order calculus. From this perspective, advanced type disciplines such as \NSTs bring substantial expressive power, beyond usual recursive types~\cite{gay2022differentshadesinfinitesession}; still, it is interesting to note that correctly encoding Minsky machines is not trivial, even when counting with the power of (nested) recursion and polymorphism in types. 

\paragraph{Acknowledgments}
We are grateful to the anonymous reviewers for their constructive feedback. 
This research has been partially supported by 
the Dutch Research Council (NWO) under project  ``\href{https://doi.org/10.61686/FHYZO53064}{Cyclic Structures in Programs and
Proofs}'' (OCENW.XL.23.089).

\bibliographystyle{eptcs}

\begin{thebibliography}{10}
\providecommand{\bibitemdeclare}[2]{}
\providecommand{\surnamestart}{}
\providecommand{\surnameend}{}
\providecommand{\urlprefix}{Available at }
\providecommand{\url}[1]{\texttt{#1}}
\providecommand{\href}[2]{\texttt{#2}}
\providecommand{\urlalt}[2]{\href{#1}{#2}}
\providecommand{\doi}[1]{doi:\urlalt{https://doi.org/#1}{#1}}
\providecommand{\eprint}[1]{arXiv:\urlalt{https://arxiv.org/abs/#1}{#1}}
\providecommand{\bibinfo}[2]{#2}

\bibitemdeclare{article}{DBLP:journals/iandc/AlmeidaMTV22}
\bibitem{DBLP:journals/iandc/AlmeidaMTV22}
\bibinfo{author}{Bernardo \surnamestart Almeida\surnameend},
  \bibinfo{author}{Andreia \surnamestart Mordido\surnameend},
  \bibinfo{author}{Peter \surnamestart Thiemann\surnameend} \&
  \bibinfo{author}{Vasco~T. \surnamestart Vasconcelos\surnameend}
  (\bibinfo{year}{2022}): \emph{\bibinfo{title}{Polymorphic lambda calculus
  with context-free session types}}.
\newblock {\slshape \bibinfo{journal}{Inf. Comput.}}
  \bibinfo{volume}{289}(\bibinfo{number}{Part}), p. \bibinfo{pages}{104948},
  \doi{10.1016/J.IC.2022.104948}.
\newblock \urlprefix\url{https://doi.org/10.1016/j.ic.2022.104948}.

\bibitemdeclare{inproceedings}{DBLP:conf/tacas/AlmeidaMV20}
\bibitem{DBLP:conf/tacas/AlmeidaMV20}
\bibinfo{author}{Bernardo \surnamestart Almeida\surnameend},
  \bibinfo{author}{Andreia \surnamestart Mordido\surnameend} \&
  \bibinfo{author}{Vasco~T. \surnamestart Vasconcelos\surnameend}
  (\bibinfo{year}{2020}): \emph{\bibinfo{title}{Deciding the Bisimilarity of
  Context-Free Session Types}}.
\newblock In \bibinfo{editor}{Armin \surnamestart Biere\surnameend} \&
  \bibinfo{editor}{David \surnamestart Parker\surnameend}, editors: {\slshape
  \bibinfo{booktitle}{Tools and Algorithms for the Construction and Analysis of
  Systems - 26th International Conference, {TACAS} 2020, Held as Part of the
  European Joint Conferences on Theory and Practice of Software, {ETAPS} 2020,
  Dublin, Ireland, April 25-30, 2020, Proceedings, Part {II}}}, {\slshape
  \bibinfo{series}{Lecture Notes in Computer Science}} \bibinfo{volume}{12079},
  \bibinfo{publisher}{Springer}, pp. \bibinfo{pages}{39--56},
  \doi{10.1007/978-3-030-45237-7\_3}.
\newblock \urlprefix\url{https://doi.org/10.1007/978-3-030-45237-7\_3}.

\bibitemdeclare{inproceedings}{DBLP:conf/concur/BalzerPT18}
\bibitem{DBLP:conf/concur/BalzerPT18}
\bibinfo{author}{Stephanie \surnamestart Balzer\surnameend},
  \bibinfo{author}{Frank \surnamestart Pfenning\surnameend} \&
  \bibinfo{author}{Bernardo \surnamestart Toninho\surnameend}
  (\bibinfo{year}{2018}): \emph{\bibinfo{title}{A Universal Session Type for
  Untyped Asynchronous Communication}}.
\newblock In \bibinfo{editor}{Sven \surnamestart Schewe\surnameend} \&
  \bibinfo{editor}{Lijun \surnamestart Zhang\surnameend}, editors: {\slshape
  \bibinfo{booktitle}{29th International Conference on Concurrency Theory,
  {CONCUR} 2018, Beijing, China, September 4-7, 2018}}, {\slshape
  \bibinfo{series}{LIPIcs}} \bibinfo{volume}{118}, \bibinfo{publisher}{Schloss
  Dagstuhl - Leibniz-Zentrum f{\"{u}}r Informatik}, pp.
  \bibinfo{pages}{30:1--30:18}, \doi{10.4230/LIPICS.CONCUR.2018.30}.
\newblock \urlprefix\url{https://doi.org/10.4230/LIPIcs.CONCUR.2018.30}.

\bibitemdeclare{inproceedings}{DBLP:conf/icalp/BusiGZ03}
\bibitem{DBLP:conf/icalp/BusiGZ03}
\bibinfo{author}{Nadia \surnamestart Busi\surnameend},
  \bibinfo{author}{Maurizio \surnamestart Gabbrielli\surnameend} \&
  \bibinfo{author}{Gianluigi \surnamestart Zavattaro\surnameend}
  (\bibinfo{year}{2003}): \emph{\bibinfo{title}{Replication vs. Recursive
  Definitions in Channel Based Calculi}}.
\newblock In \bibinfo{editor}{Jos C.~M. \surnamestart Baeten\surnameend},
  \bibinfo{editor}{Jan~Karel \surnamestart Lenstra\surnameend},
  \bibinfo{editor}{Joachim \surnamestart Parrow\surnameend} \&
  \bibinfo{editor}{Gerhard~J. \surnamestart Woeginger\surnameend}, editors:
  {\slshape \bibinfo{booktitle}{Automata, Languages and Programming, 30th
  International Colloquium, {ICALP} 2003, Eindhoven, The Netherlands, June 30 -
  July 4, 2003. Proceedings}}, {\slshape \bibinfo{series}{Lecture Notes in
  Computer Science}} \bibinfo{volume}{2719}, \bibinfo{publisher}{Springer}, pp.
  \bibinfo{pages}{133--144}, \doi{10.1007/3-540-45061-0\_12}.
\newblock \urlprefix\url{https://doi.org/10.1007/3-540-45061-0\_12}.

\bibitemdeclare{inproceedings}{DBLP:conf/icalp/BusiGZ04}
\bibitem{DBLP:conf/icalp/BusiGZ04}
\bibinfo{author}{Nadia \surnamestart Busi\surnameend},
  \bibinfo{author}{Maurizio \surnamestart Gabbrielli\surnameend} \&
  \bibinfo{author}{Gianluigi \surnamestart Zavattaro\surnameend}
  (\bibinfo{year}{2004}): \emph{\bibinfo{title}{Comparing Recursion,
  Replication, and Iteration in Process Calculi}}.
\newblock In \bibinfo{editor}{Josep \surnamestart D{\'{\i}}az\surnameend},
  \bibinfo{editor}{Juhani \surnamestart Karhum{\"{a}}ki\surnameend},
  \bibinfo{editor}{Arto \surnamestart Lepist{\"{o}}\surnameend} \&
  \bibinfo{editor}{Donald \surnamestart Sannella\surnameend}, editors:
  {\slshape \bibinfo{booktitle}{Automata, Languages and Programming: 31st
  International Colloquium, {ICALP} 2004, Turku, Finland, July 12-16, 2004.
  Proceedings}}, {\slshape \bibinfo{series}{Lecture Notes in Computer Science}}
  \bibinfo{volume}{3142}, \bibinfo{publisher}{Springer}, pp.
  \bibinfo{pages}{307--319}, \doi{10.1007/978-3-540-27836-8\_28}.
\newblock \urlprefix\url{https://doi.org/10.1007/978-3-540-27836-8\_28}.

\bibitemdeclare{article}{DBLP:journals/mscs/BusiGZ09}
\bibitem{DBLP:journals/mscs/BusiGZ09}
\bibinfo{author}{Nadia \surnamestart Busi\surnameend},
  \bibinfo{author}{Maurizio \surnamestart Gabbrielli\surnameend} \&
  \bibinfo{author}{Gianluigi \surnamestart Zavattaro\surnameend}
  (\bibinfo{year}{2009}): \emph{\bibinfo{title}{On the expressive power of
  recursion, replication and iteration in process calculi}}.
\newblock {\slshape \bibinfo{journal}{Math. Struct. Comput. Sci.}}
  \bibinfo{volume}{19}(\bibinfo{number}{6}), pp. \bibinfo{pages}{1191--1222},
  \doi{10.1017/S096012950999017X}.
\newblock \urlprefix\url{https://doi.org/10.1017/S096012950999017X}.

\bibitemdeclare{inproceedings}{DBLP:journals/corr/abs-2505-20848}
\bibitem{DBLP:journals/corr/abs-2505-20848}
\bibinfo{author}{Lu{\'{\i}}s \surnamestart Caires\surnameend}
  (\bibinfo{year}{2025}): \emph{\bibinfo{title}{Thread and Memory-Safe
  Programming with {CLASS}}}.
\newblock In \bibinfo{editor}{Farzaneh \surnamestart Derakhshan\surnameend} \&
  \bibinfo{editor}{Jan \surnamestart Hoffmann\surnameend}, editors: {\slshape
  \bibinfo{booktitle}{Proceedings 16th International Workshop on Programming
  Language Approaches to Concurrency and Communication-cEntric Software,
  PLACES@ETAPS 2025, Hamilton, Canada, 4th May 2025}}, {\slshape
  \bibinfo{series}{{EPTCS}}} \bibinfo{volume}{420}, pp.
  \bibinfo{pages}{22--33}, \doi{10.4204/EPTCS.420.3}.
\newblock \urlprefix\url{https://doi.org/10.4204/EPTCS.420.3}.

\bibitemdeclare{inproceedings}{DBLP:conf/concur/CairesP10}
\bibitem{DBLP:conf/concur/CairesP10}
\bibinfo{author}{Lu{\'{\i}}s \surnamestart Caires\surnameend} \&
  \bibinfo{author}{Frank \surnamestart Pfenning\surnameend}
  (\bibinfo{year}{2010}): \emph{\bibinfo{title}{Session Types as Intuitionistic
  Linear Propositions}}.
\newblock In \bibinfo{editor}{Paul \surnamestart Gastin\surnameend} \&
  \bibinfo{editor}{Fran{\c{c}}ois \surnamestart Laroussinie\surnameend},
  editors: {\slshape \bibinfo{booktitle}{{CONCUR} 2010 - Concurrency Theory,
  21th International Conference, {CONCUR} 2010, Paris, France, August
  31-September 3, 2010. Proceedings}}, {\slshape \bibinfo{series}{Lecture Notes
  in Computer Science}} \bibinfo{volume}{6269}, \bibinfo{publisher}{Springer},
  pp. \bibinfo{pages}{222--236}, \doi{10.1007/978-3-642-15375-4\_16}.
\newblock \urlprefix\url{https://doi.org/10.1007/978-3-642-15375-4\_16}.

\bibitemdeclare{inproceedings}{DBLP:conf/esop/DasDMP21}
\bibitem{DBLP:conf/esop/DasDMP21}
\bibinfo{author}{Ankush \surnamestart Das\surnameend}, \bibinfo{author}{Henry
  \surnamestart DeYoung\surnameend}, \bibinfo{author}{Andreia \surnamestart
  Mordido\surnameend} \& \bibinfo{author}{Frank \surnamestart
  Pfenning\surnameend} (\bibinfo{year}{2021}): \emph{\bibinfo{title}{Nested
  Session Types}}.
\newblock In \bibinfo{editor}{Nobuko \surnamestart Yoshida\surnameend}, editor:
  {\slshape \bibinfo{booktitle}{Programming Languages and Systems - 30th
  European Symposium on Programming, {ESOP} 2021, Held as Part of the European
  Joint Conferences on Theory and Practice of Software, {ETAPS} 2021,
  Luxembourg City, Luxembourg, March 27 - April 1, 2021, Proceedings}},
  {\slshape \bibinfo{series}{Lecture Notes in Computer Science}}
  \bibinfo{volume}{12648}, \bibinfo{publisher}{Springer}, pp.
  \bibinfo{pages}{178--206}, \doi{10.1007/978-3-030-72019-3\_7}.
\newblock \urlprefix\url{https://doi.org/10.1007/978-3-030-72019-3\_7}.

\bibitemdeclare{article}{das2020nestedsessiontypes}
\bibitem{das2020nestedsessiontypes}
\bibinfo{author}{Ankush \surnamestart Das\surnameend}, \bibinfo{author}{Henry
  \surnamestart DeYoung\surnameend}, \bibinfo{author}{Andreia \surnamestart
  Mordido\surnameend} \& \bibinfo{author}{Frank \surnamestart
  Pfenning\surnameend} (\bibinfo{year}{2022}): \emph{\bibinfo{title}{Nested
  Session Types}}.
\newblock {\slshape \bibinfo{journal}{{ACM} Trans. Program. Lang. Syst.}}
  \bibinfo{volume}{44}(\bibinfo{number}{3}), pp. \bibinfo{pages}{19:1--19:45},
  \doi{10.1145/3539656}.
\newblock \urlprefix\url{https://doi.org/10.1145/3539656}.

\bibitemdeclare{inproceedings}{gay2022differentshadesinfinitesession}
\bibitem{gay2022differentshadesinfinitesession}
\bibinfo{author}{Simon~J. \surnamestart Gay\surnameend}, \bibinfo{author}{Diogo
  \surnamestart Po{\c{c}}as\surnameend} \& \bibinfo{author}{Vasco~T.
  \surnamestart Vasconcelos\surnameend} (\bibinfo{year}{2022}):
  \emph{\bibinfo{title}{The Different Shades of Infinite Session Types}}.
\newblock In \bibinfo{editor}{Patricia \surnamestart Bouyer\surnameend} \&
  \bibinfo{editor}{Lutz \surnamestart Schr{\"{o}}der\surnameend}, editors:
  {\slshape \bibinfo{booktitle}{Foundations of Software Science and Computation
  Structures - 25th International Conference, {FOSSACS} 2022, Held as Part of
  the European Joint Conferences on Theory and Practice of Software, {ETAPS}
  2022, Munich, Germany, April 2-7, 2022, Proceedings}}, {\slshape
  \bibinfo{series}{Lecture Notes in Computer Science}} \bibinfo{volume}{13242},
  \bibinfo{publisher}{Springer}, pp. \bibinfo{pages}{347--367},
  \doi{10.1007/978-3-030-99253-8\_18}.
\newblock \urlprefix\url{https://doi.org/10.1007/978-3-030-99253-8\_18}.

\bibitemdeclare{inproceedings}{DBLP:conf/concur/Honda93}
\bibitem{DBLP:conf/concur/Honda93}
\bibinfo{author}{Kohei \surnamestart Honda\surnameend} (\bibinfo{year}{1993}):
  \emph{\bibinfo{title}{Types for Dyadic Interaction}}.
\newblock In \bibinfo{editor}{Eike \surnamestart Best\surnameend}, editor:
  {\slshape \bibinfo{booktitle}{{CONCUR} '93, 4th International Conference on
  Concurrency Theory, Hildesheim, Germany, August 23-26, 1993, Proceedings}},
  {\slshape \bibinfo{series}{Lecture Notes in Computer Science}}
  \bibinfo{volume}{715}, \bibinfo{publisher}{Springer}, pp.
  \bibinfo{pages}{509--523}, \doi{10.1007/3-540-57208-2\_35}.
\newblock \urlprefix\url{https://doi.org/10.1007/3-540-57208-2\_35}.

\bibitemdeclare{inproceedings}{DBLP:conf/esop/HondaVK98}
\bibitem{DBLP:conf/esop/HondaVK98}
\bibinfo{author}{Kohei \surnamestart Honda\surnameend},
  \bibinfo{author}{Vasco~Thudichum \surnamestart Vasconcelos\surnameend} \&
  \bibinfo{author}{Makoto \surnamestart Kubo\surnameend}
  (\bibinfo{year}{1998}): \emph{\bibinfo{title}{Language Primitives and Type
  Discipline for Structured Communication-Based Programming}}.
\newblock In \bibinfo{editor}{Chris \surnamestart Hankin\surnameend}, editor:
  {\slshape \bibinfo{booktitle}{Programming Languages and Systems - ESOP'98,
  7th European Symposium on Programming, Held as Part of the European Joint
  Conferences on the Theory and Practice of Software, ETAPS'98, Lisbon,
  Portugal, March 28 - April 4, 1998, Proceedings}}, {\slshape
  \bibinfo{series}{Lecture Notes in Computer Science}} \bibinfo{volume}{1381},
  \bibinfo{publisher}{Springer}, pp. \bibinfo{pages}{122--138},
  \doi{10.1007/BFB0053567}.
\newblock \urlprefix\url{https://doi.org/10.1007/BFb0053567}.

\bibitemdeclare{misc}{rfc4976}
\bibitem{rfc4976}
\bibinfo{author}{C.~\surnamestart Jennings\surnameend},
  \bibinfo{author}{R.~\surnamestart Mahy\surnameend} \& \bibinfo{author}{A.~B.
  \surnamestart Roach\surnameend} (\bibinfo{year}{2007}):
  \emph{\bibinfo{title}{{Relay Extensions for the Message Sessions Relay
  Protocol (MSRP)}}}.
\newblock \bibinfo{howpublished}{RFC 4976}, \doi{10.17487/RFC4976}.
\newblock \urlprefix\url{https://www.rfc-editor.org/info/rfc4976/}.

\bibitemdeclare{inproceedings}{DBLP:conf/esop/KouzapasPY16}
\bibitem{DBLP:conf/esop/KouzapasPY16}
\bibinfo{author}{Dimitrios \surnamestart Kouzapas\surnameend},
  \bibinfo{author}{Jorge~A. \surnamestart P{\'{e}}rez\surnameend} \&
  \bibinfo{author}{Nobuko \surnamestart Yoshida\surnameend}
  (\bibinfo{year}{2016}): \emph{\bibinfo{title}{On the Relative Expressiveness
  of Higher-Order Session Processes}}.
\newblock In \bibinfo{editor}{Peter \surnamestart Thiemann\surnameend}, editor:
  {\slshape \bibinfo{booktitle}{Programming Languages and Systems - 25th
  European Symposium on Programming, {ESOP} 2016, Held as Part of the European
  Joint Conferences on Theory and Practice of Software, {ETAPS} 2016,
  Eindhoven, The Netherlands, April 2-8, 2016, Proceedings}}, {\slshape
  \bibinfo{series}{Lecture Notes in Computer Science}} \bibinfo{volume}{9632},
  \bibinfo{publisher}{Springer}, pp. \bibinfo{pages}{446--475},
  \doi{10.1007/978-3-662-49498-1\_18}.
\newblock \urlprefix\url{https://doi.org/10.1007/978-3-662-49498-1\_18}.

\bibitemdeclare{article}{DBLP:journals/iandc/KouzapasPY19}
\bibitem{DBLP:journals/iandc/KouzapasPY19}
\bibinfo{author}{Dimitrios \surnamestart Kouzapas\surnameend},
  \bibinfo{author}{Jorge~A. \surnamestart P{\'{e}}rez\surnameend} \&
  \bibinfo{author}{Nobuko \surnamestart Yoshida\surnameend}
  (\bibinfo{year}{2019}): \emph{\bibinfo{title}{On the relative expressiveness
  of higher-order session processes}}.
\newblock {\slshape \bibinfo{journal}{Inf. Comput.}} \bibinfo{volume}{268},
  \doi{10.1016/J.IC.2019.06.002}.
\newblock \urlprefix\url{https://doi.org/10.1016/j.ic.2019.06.002}.

\bibitemdeclare{inproceedings}{DBLP:conf/lics/LanesePSS08}
\bibitem{DBLP:conf/lics/LanesePSS08}
\bibinfo{author}{Ivan \surnamestart Lanese\surnameend},
  \bibinfo{author}{Jorge~A. \surnamestart P{\'{e}}rez\surnameend},
  \bibinfo{author}{Davide \surnamestart Sangiorgi\surnameend} \&
  \bibinfo{author}{Alan \surnamestart Schmitt\surnameend}
  (\bibinfo{year}{2008}): \emph{\bibinfo{title}{On the Expressiveness and
  Decidability of Higher-Order Process Calculi}}.
\newblock In: {\slshape \bibinfo{booktitle}{Proceedings of the Twenty-Third
  Annual {IEEE} Symposium on Logic in Computer Science, {LICS} 2008, 24-27 June
  2008, Pittsburgh, PA, {USA}}}, \bibinfo{publisher}{{IEEE} Computer Society},
  pp. \bibinfo{pages}{145--155}, \doi{10.1109/LICS.2008.8}.
\newblock \urlprefix\url{https://doi.org/10.1109/LICS.2008.8}.

\bibitemdeclare{article}{DBLP:journals/iandc/LanesePSS11}
\bibitem{DBLP:journals/iandc/LanesePSS11}
\bibinfo{author}{Ivan \surnamestart Lanese\surnameend},
  \bibinfo{author}{Jorge~A. \surnamestart P{\'{e}}rez\surnameend},
  \bibinfo{author}{Davide \surnamestart Sangiorgi\surnameend} \&
  \bibinfo{author}{Alan \surnamestart Schmitt\surnameend}
  (\bibinfo{year}{2011}): \emph{\bibinfo{title}{On the expressiveness and
  decidability of higher-order process calculi}}.
\newblock {\slshape \bibinfo{journal}{Inf. Comput.}}
  \bibinfo{volume}{209}(\bibinfo{number}{2}), pp. \bibinfo{pages}{198--226},
  \doi{10.1016/J.IC.2010.10.001}.
\newblock \urlprefix\url{https://doi.org/10.1016/j.ic.2010.10.001}.

\bibitemdeclare{book}{minsky1967computation}
\bibitem{minsky1967computation}
\bibinfo{author}{Marvin~L. \surnamestart Minsky\surnameend}
  (\bibinfo{year}{1967}): \emph{\bibinfo{title}{Computation: finite and
  infinite machines}}.
\newblock \bibinfo{publisher}{Prentice-Hall, Inc.}, \bibinfo{address}{USA}.

\bibitemdeclare{inproceedings}{DBLP:conf/esop/RochaC23}
\bibitem{DBLP:conf/esop/RochaC23}
\bibinfo{author}{Pedro \surnamestart Rocha\surnameend} \&
  \bibinfo{author}{Lu{\'{\i}}s \surnamestart Caires\surnameend}
  (\bibinfo{year}{2023}): \emph{\bibinfo{title}{Safe Session-Based Concurrency
  with Shared Linear State}}.
\newblock In \bibinfo{editor}{Thomas \surnamestart Wies\surnameend}, editor:
  {\slshape \bibinfo{booktitle}{Programming Languages and Systems - 32nd
  European Symposium on Programming, {ESOP} 2023, Held as Part of the European
  Joint Conferences on Theory and Practice of Software, {ETAPS} 2023, Paris,
  France, April 22-27, 2023, Proceedings}}, {\slshape \bibinfo{series}{Lecture
  Notes in Computer Science}} \bibinfo{volume}{13990},
  \bibinfo{publisher}{Springer}, pp. \bibinfo{pages}{421--450},
  \doi{10.1007/978-3-031-30044-8\_16}.
\newblock \urlprefix\url{https://doi.org/10.1007/978-3-031-30044-8\_16}.

\bibitemdeclare{inproceedings}{DBLP:conf/concur/SilvaMV23}
\bibitem{DBLP:conf/concur/SilvaMV23}
\bibinfo{author}{Gil \surnamestart Silva\surnameend}, \bibinfo{author}{Andreia
  \surnamestart Mordido\surnameend} \& \bibinfo{author}{Vasco~T. \surnamestart
  Vasconcelos\surnameend} (\bibinfo{year}{2023}):
  \emph{\bibinfo{title}{Subtyping Context-Free Session Types}}.
\newblock In \bibinfo{editor}{Guillermo~A. \surnamestart
  P{\'{e}}rez\surnameend} \& \bibinfo{editor}{Jean{-}Fran{\c{c}}ois
  \surnamestart Raskin\surnameend}, editors: {\slshape \bibinfo{booktitle}{34th
  International Conference on Concurrency Theory, {CONCUR} 2023, Antwerp,
  Belgium, September 18-23, 2023}}, {\slshape \bibinfo{series}{LIPIcs}}
  \bibinfo{volume}{279}, \bibinfo{publisher}{Schloss Dagstuhl - Leibniz-Zentrum
  f{\"{u}}r Informatik}, pp. \bibinfo{pages}{11:1--11:19},
  \doi{10.4230/LIPICS.CONCUR.2023.11}.
\newblock \urlprefix\url{https://doi.org/10.4230/LIPIcs.CONCUR.2023.11}.

\bibitemdeclare{inproceedings}{DBLP:conf/icfp/ThiemannV16}
\bibitem{DBLP:conf/icfp/ThiemannV16}
\bibinfo{author}{Peter \surnamestart Thiemann\surnameend} \&
  \bibinfo{author}{Vasco~T. \surnamestart Vasconcelos\surnameend}
  (\bibinfo{year}{2016}): \emph{\bibinfo{title}{Context-free session types}}.
\newblock In \bibinfo{editor}{Jacques \surnamestart Garrigue\surnameend},
  \bibinfo{editor}{Gabriele \surnamestart Keller\surnameend} \&
  \bibinfo{editor}{Eijiro \surnamestart Sumii\surnameend}, editors: {\slshape
  \bibinfo{booktitle}{Proceedings of the 21st {ACM} {SIGPLAN} International
  Conference on Functional Programming, {ICFP} 2016, Nara, Japan, September
  18-22, 2016}}, \bibinfo{publisher}{{ACM}}, pp. \bibinfo{pages}{462--475},
  \doi{10.1145/2951913.2951926}.
\newblock \urlprefix\url{https://doi.org/10.1145/2951913.2951926}.

\bibitemdeclare{mastersthesis}{sjors2026}
\bibitem{sjors2026}
\bibinfo{author}{Sjors \surnamestart Wortelboer\surnameend}
  (\bibinfo{year}{2026}): \emph{\bibinfo{title}{Nested Session Types in Turing
  Complete Calculi}}.
\newblock \bibinfo{type}{Master's thesis}, \bibinfo{school}{University of
  Groningen}.
\newblock \urlprefix\url{https://jperez.nl/files/MScThesis-Wortelboer.pdf}.
\newblock \bibinfo{note}{Accessed: 2026-06-28}.

\end{thebibliography}

\iftoggle{extended}{

\newpage
\appendix

\section{Additional Material for \Cref{s:back}}

\subsection{Type rules \pNSTs (\Cref{def:type-rules-pst})}
\label{apx:type-rules-pst}

Type rules for processes $P$:
\begin{mathpar}
\AXC{$(k\in L)$}
        \AXC{$\nV;\Delta\nT P :: (x:A_k)$}
    \RightLabel{$\oplus R$}
    \BIC{$\nV;\Delta\nT \nChoose{x}{k} P :: (x:\ntChoose{\nOpt{l} A_l}_{l\in L})$}
    \DisplayProof
    
        \AXC{$(\forall l\in L)$}
        \AXC{$\nV;\Delta,x:A_l\nT Q_l :: (z:C)$}
    \RightLabel{$\oplus L$}
    \BIC{$\nV;\Delta,x:\ntChoose{\ntOpt{l} A_l}_{l\in L}\nT \nCase{x}[\nOpt{l} Q_l]_{l\in L} :: (z:C)$}
    \DisplayProof

\AXC{$(\forall l\in L)$}
        \AXC{$\nV;\Delta\nT P_l :: (x:A_l)$}
    \RightLabel{$\& R$}
    \BIC{$\nV;\Delta\nT \nCase{x}[\nOpt{l} Q_l]_{l\in L} :: (x:\ntOffer{\nOpt{l} A_l}_{l\in L})$}
    \DisplayProof

        \AXC{$(k\in L)$}
        \AXC{$\nV;\Delta\nT P :: (x:A_k)$}
    \RightLabel{$\& L$}
    \BIC{$\nV;\Delta,x:\ntOffer{\nOpt{l} A_l}_{l\in L}\nT \nChoose{x}{k} Q :: (z:C)$}
    \DisplayProof

\AXC{$\cdot;\nV\nT A\equiv A'$}
        \AXC{$\nV;\Delta\nT P :: (x:B)$}
    \RightLabel{$\otimes R$}
    \BIC{$\nV;\Delta,y:A\nT \nSend{x}{y}P :: (x:A'\ntSend B)$}
    \DisplayProof

        \AXC{$\nV;\Delta,y:A,x:B\nT Q :: (z:C)$}
    \RightLabel{$\otimes L$}
    \UIC{$\nV;\Delta,x:A\ntSend B\nT \nRecv{y}{x}Q :: (z:C)$}
    \DisplayProof

\AXC{$\nV;\Delta,y:A\nT P :: (x:B)$}
    \RightLabel{$\multimap\! R$}
    \UIC{$\nV;\Delta\nT \nRecv{y}{x}P :: (x:A\ntRecv B)$}
    \DisplayProof

        \AXC{$\cdot;\nV\nT A\equiv A'$}
        \AXC{$\nV;\Delta,x:B\nT Q :: (z:C)$}
    \RightLabel{$\multimap\! L$}
    \BIC{$\nV;\Delta,x:A'\ntRecv B,y:A\nT \nSend{x}{y}Q :: (z:C)$}
    \DisplayProof

\AXC{}
    \RightLabel{$\ntOne R$}
    \UIC{$\cdot;\cdot\nT \nClose{x} :: (x:\ntOne)$}
    \DisplayProof

        \AXC{$\nV;\Delta\nT Q :: (z:C)$}
    \RightLabel{$\ntOne L$}
    \UIC{$\nV;\Delta,x:\ntOne\nT \nWait{x} Q :: (z:C)$}
    \DisplayProof

\AXC{$\cdot;\nV\nT A\equiv A'$}
    \RightLabel{id}
    \UIC{$\nV;y:A\nT \nFwd{x}{y} :: (x:A')$}
    \DisplayProof

\AXC{$\overline{y':B'}\nT f[\overline{\nAlpha}] = P_f  :: (x':B) \in \Sigma$}
        \AXC{$\cdot;\nV\nT\Delta'\equiv \overline{y:B'[\nTheta]}$}
        \AXC{$\nV;\Delta,x:B[\nTheta]\nT Q :: (z:C)$}
    \RightLabel{def}
    \TIC{$\nV;\Delta,\Delta'\nT \nPd{x}{f}[\nTheta]<\overline{y}> Q :: (z:C)$}
    \DisplayProof
\end{mathpar}

Type rules for messages $M$:
\begin{mathpar}
\AXC{$(k\in L)$}
    \RightLabel{$\oplus M$}
    \UIC{$\cdot; x':A_k \nT \nChoose{x}{k} \nFwd{x}{x'} :: (x:\ntChoose{\ntOpt{l} A_l}_{l\in L})$}
    \DisplayProof

\AXC{$(k\in L)$}
    \RightLabel{$\& M$}
    \UIC{$\cdot; x:\ntOffer{\ntOpt{l} A_l}_{l\in L} \nT \nChoose{x}{k} \nFwd{x'}{x} :: (x':A_k)$}
    \DisplayProof

\AXC{$\cdot;\cdot\nT A\equiv A'$}
    \RightLabel{$\otimes M$}
    \UIC{$\cdot; y: A, x': B \nT \nSend{x}{y} \nFwd{x}{x'} :: (x:A'\ntSend B)$}
    \DisplayProof

\AXC{$\cdot;\cdot\nT A\equiv A'$}
    \RightLabel{$\multimap\! M$}
    \UIC{$\cdot; x: A'\ntRecv B, y:A \nT \nSend{x}{y} \nFwd{x'}{x} :: (x':B)$}
    \DisplayProof

\AXC{}
    \RightLabel{$\ntOne M$}
    \UIC{$\cdot; \cdot \nT \nClose{x} :: (x:\ntOne)$}
    \DisplayProof
\end{mathpar}

\subsection{Reduction rules for \pNSTs (\Cref{def:reductions-pst})}
\label{apx:reductions-pst}

\begin{align*}
    &\nProc{x}{\nChoose{x}{k} P}
        \pred \nProc{x'}{P[x'/x]}, \nmLabel{x}{x'}{k}
        & (\oplus S) \\*
    &\nmLabel{x}{x'}{k}, \nProc{y}{\nCase{x}[\nOpt{l} Q_l]}
        \pred \nProc{y}{Q_k[x'/x]}
        & (\oplus C) \\*
    \\
    &\nProc{d}{\nChoose{x}{k} P}
        \pred\nmLabel{x}{x'}{k}, \nProc{d}{P[x'/x]}
        & (\& S) \\*
    &\nProc{x}{\nCase{x}[\nOpt{l} Q_l]}, \nmLabel{x}{x'}{k}
        \pred \nProc{x'}{Q_k[x'/x]} 
        & (\& C) \\*
    \\
    &\nProc{x}{\nSend{x}{w} P}
        \pred \nProc{x'}{P[x'/x]}, \nmSend{x}{x'}{w}
        & (\otimes S) \\*
    &\nmSend{x}{x'}{w}, \nProc{y}{\nRecv{w'}{x} Q}
        \pred \nProc{y}{Q[x'/x,w/w']}
        & (\otimes C) \\*
    \\
    &\nProc{y}{\nSend{x}{w} P}
        \pred \nmSend{x}{x'}{w}, \nProc{y}{P[x'/x]}
        & (\multimap S) \\*
    &\nProc{x}{\nRecv{w'}{x} Q}, \nmSend{x}{x'}{w}
        \pred \nProc{x'}{Q[x'/x,w/w']}
        & (\multimap C) \\*
    \\
    &\nProc{x}{\nClose{x}}
        \pred \nmClose{x}
        & (\ntOne S) \\*
    &\nmClose{x}, \nProc{y}{\nWait{x} P}
        \pred \nProc{y}{P}
        & (\ntOne C) \\*
    \\
    &\nMsg{d}{M},\nProc{c}{\nFwd{c}{d}}
        \pred \nMsg{c}{M[c/d]}
        & (\text{id$^+$C}) \\*
    &\nProc{c}{\nFwd{c}{d}},\nMsg{e}{M_c}
        \pred \nMsg{e}{M_c[d/c]}
        & (\text{id$^-$C}) \\*
    \\
    &\nProc{d}{Q},\nProc{c}{\nFwd{c}{d}}
        \pred \nProc{c}{Q[c/d]}
        & (\text{id$^+$S}) \\*
    &\nProc{c}{\nFwd{c}{d}},\nProc{e}{Q_c}
        \pred \nProc{e}{Q_c[d/c]}
        & (\text{id$^-$S}) \\*
    \\
    &\nProc{z}{\nPd{x}{f}[\nTheta]<\overline{d}> Q}
        \pred \nProc{a}{P_f[a/x,\overline{d}/\overline{y}, \nTheta]},
        \nProc{z}{Q[a/x]}
        & (\text{defC})
\end{align*}

\section{Additional Material for \Cref{s:encoding}}

\subsection{Constituent process definitions (\Cref{def:instantiation-encoding})}
\label{apx:instantiation-nodes}

\begin{align*}
        \begin{split}
N^{0,0} = \nCase{z}
                &\nOpt{exp_0} \nPd{c}{t^{0,0}} \nPdF{z}{t^{1,0}}<c> \\
                &\nOpt{exp_1} \nPd{c}{t^{0,0}} \nPdF{z}{t^{0,1}}<c> \\
                &\nOpt{dup} \nPd{c}{t^{0,0}} \nSend{z}{c} \nPdF{z}{t^{0,0}} \\
                &\nOpt{ter} \nClose{z}
            \nEndCase
        \end{split} \\
        \begin{split}
N^{1,0} = \nCase{z}
                &\nOpt{ins_0} \nRecv{r_0}{z} \nRecv{r_1}{z} \nRecv{y}{z} \\
                    &\qquad\nCase{z}
                        \nOpt{i} \nPdF{z}{p^{0,0}_i}<r_0~r_1~a~y>
                    \nEndCase_{i\in L\mM} \\
                &\nOpt{exp_0} \nChoose{a}{exp_0} \nPdF{z}{t^{s,0}}[0[]]<a> \\
                &\nOpt{exp_1} \nChoose{a}{dup}
                                    \nRecv{b}{a}
                                    \nChoose{a}{exp_1}
                                    \nChoose{b}{exp_0} 
                                    \nPdF{z}{t^{1,1}}<a~b>\\
                &\nOpt{dup} \nChoose{a}{dup} 
                                    \nRecv{a'}{a}
                                    \nPd{c}{t^{1,0}}<a>
                                    \nSend{z}{c}
                                    \nPdF{z}{t^{1,0}}<a'> \\
                &\nOpt{ter} \nChoose{a}{1} \nWait{a} \nClose{z}
            \nEndCase
        \end{split} \\
        \begin{split}
N^{s,0} = \nCase{z}
                &\nOpt{ins_0} \nRecv{r_0}{z} \nRecv{r_1}{z} \nRecv{y}{z} \\
                    &\qquad\nCase{z}
                        \nOpt{i} \nPdF{z}{p^{s,0}_i}[\nAlpha]<r_0~r_1~a~y>
                    \nEndCase_{i\in L\mM} \\
                &\nOpt{exp_0} \nChoose{a}{exp_0} \nPdF{z}{t^{s,0}}[S[\nAlpha]]<a> \\
                &\nOpt{exp_1} \nChoose{a}{dup}
                                    \nRecv{b}{a}
                                    \nChoose{a}{exp_1}
                                    \nChoose{b}{exp_0} 
                                    \nPdF{z}{t^{s,1}}[\nAlpha]<a~b> \\
                &\nOpt{dup} \nChoose{a}{dup}
                                    \nRecv{b}{a}
                                    \nPd{c}{t^{s,0}}[\nAlpha]<a>
                                    \nSend{z}{c}
                                    \nPdF{z}{t^{s,0}}[\nAlpha]<b> \\
                &\nOpt{ter} \nChoose{a}{ter} \nWait{a} \nClose{z}
            \nEndCase
        \end{split} \\
        \begin{split}
N^{0,1} = \nCase{z}
                &\nOpt{ins_1} \nRecv{r_0}{z} \nRecv{r_1}{z} \nRecv{y}{z} \\
                    &\qquad\nCase{z}
                    \nOpt{i} \nPdF{z}{p^{0,0}_i}<r_0~r_1~b~y>
                    \nEndCase_{i\in L\mM} \\
                &\nOpt{exp_0} \nChoose{b}{dup}
                                    \nRecv{a}{b}
                                    \nChoose{a}{exp_1}
                                    \nChoose{b}{exp_0} 
                                    \nPdF{z}{t^{1,1}}<a~b>\\
                &\nOpt{exp_1} \nChoose{b}{exp_1} \nPdF{z}{t^{0,s}}[0[]]<b> \\
                &\nOpt{dup} \nChoose{a}{dup} 
                                    \nRecv{a'}{a}
                                    \nPd{c}{t^{0,1}}<a>
                                    \nSend{z}{c}
                                    \nPdF{z}{t^{0,1}}<a'> \\
                &\nOpt{ter} \nChoose{b}{1} \nWait{b} \nClose{z}
            \nEndCase
        \end{split} \\
        \begin{split}
N^{0,s} = \nCase{z}
                &\nOpt{ins_1} \nRecv{r_0}{z} \nRecv{r_1}{z} \nRecv{y}{z} \\
                    &\qquad\nCase{z}
                        \nOpt{i} \nPdF{z}{p^{s,0}_i}[\nAlpha]<r_0~r_1~a~y>
                    \nEndCase_{i\in L\mM} \\
                &\nOpt{exp_0} \nChoose{b}{dup}
                                    \nRecv{a}{b}
                                    \nChoose{b}{exp_1}
                                    \nChoose{a}{exp_0}
                                    \nPdF{z}{t^{1,s}}[\nBeta]<b~a> \\
                &\nOpt{exp_1} \nChoose{b}{exp_1} \nPdF{z}{t^{0,s}}[S[\nBeta]]<a> \\
                &\nOpt{dup} \nChoose{b}{dup}
                                    \nRecv{a}{b}
                                    \nPd{c}{t^{0,s}}[\nBeta]<b>
                                    \nSend{z}{c}
                                    \nPdF{z}{t^{0,s}}[\nBeta]<a> \\
                &\nOpt{ter} \nChoose{b}{ter} \nWait{b} \nClose{z}
            \nEndCase
        \end{split}
        \\
\begin{split}
N^{1,1} = \nCase{z}
                &\nOpt{ins_0} \nChoose{b}{ter} \nWait{b}
                    \nRecv{r_0}{z} \nRecv{r_1}{z} \nRecv{y}{z} \\
                    &\qquad\nCase{z}
                        \nOpt{i} \nPdF{z}{p^{0,s}_i}[\nAlpha]<r_0~r_1~a~y>
                    \nEndCase_{i\in L\mM} \\
                &\nOpt{ins_1} \nChoose{a}{ter} \nWait{a}
                    \nRecv{r_0}{z} \nRecv{r_1}{z} \nRecv{y}{z} \\
                    &\qquad\nCase{z}
                        \nOpt{i} \nPdF{z}{p^{s,0}_i}[\nAlpha]<r_0~r_1~b~y>
                    \nEndCase_{i\in L\mM} \\
                &\nOpt{exp_0} \nChoose{a}{exp_0} \nChoose{b}{exp_0} \nPdF{z}{t^{s,1}}[0[]]<a~b> \\
                &\nOpt{exp_1} \nChoose{a}{exp_1} \nChoose{b}{exp_1} \nPdF{z}{t^{1,s}}[0[]]<a~b> \\
                &\nOpt{ter} \nChoose{a}{ter} \nChoose{b}{ter} \nWait{a} \nWait{b} \nClose{z}
            \nEndCase
        \end{split} \\
        \begin{split}
N^{s,1} = \nCase{z}
                &\nOpt{ins_0} \nChoose{b}{ter} \nWait{b}
                    \nRecv{r_0}{z} \nRecv{r_1}{z} \nRecv{y}{z} \\
                    &\qquad\nCase{z}
                        \nOpt{i} \nPdF{z}{p^{s,s}_i}[\nAlpha]<r_0~r_1~a~y>
                    \nEndCase_{i\in L\mM} \\
                &\nOpt{ins_1} \nChoose{a}{ter} \nWait{a}
                    \nRecv{r_0}{z} \nRecv{r_1}{z} \nRecv{y}{z} \\
                    &\qquad\nCase{z}
                        \nOpt{i} \nPdF{z}{p^{s,0}_i}[\nAlpha]<r_0~r_1~b~y>
                    \nEndCase_{i\in L\mM} \\
                &\nOpt{exp_0} \nChoose{a}{exp_0} \nChoose{b}{exp_0} \nPdF{z}{t^{s,1}}[S[\nAlpha]]<a~b> \\
                &\nOpt{exp_1} \nChoose{a}{exp_1} \nChoose{b}{exp_1} \nPdF{z}{t^{s,s}}[\nAlpha,0[]]<a~b> \\
                &\nOpt{ter} \nChoose{a}{ter} \nChoose{b}{ter} \nWait{a} \nWait{b} \nClose{z}
            \nEndCase
        \end{split} \\
        \begin{split}
N^{1,s} = \nCase{z}
                &\nOpt{ins_0} \nChoose{b}{ter} \nWait{b}
                    \nRecv{r_0}{z} \nRecv{r_1}{z} \nRecv{y}{z} \\
                    &\qquad\nCase{z}
                        \nOpt{i} \nPdF{z}{p^{0,s}_i}[\nBeta]<r_0~r_1~a~y>
                    \nEndCase_{i\in L\mM} \\
                &\nOpt{ins_1} \nChoose{a}{ter} \nWait{a}
                    \nRecv{r_0}{z} \nRecv{r_1}{z} \nRecv{y}{z} \\
                    &\qquad\nCase{z}
                        \nOpt{i} \nPdF{z}{p^{s,s}_i}[\nBeta]<r_0~r_1~b~y>
                    \nEndCase_{i\in L\mM} \\
                &\nOpt{exp_0} \nChoose{a}{exp_0} \nChoose{b}{exp_0} \nPdF{z}{t^{s,s}}[0[],\nBeta]<a~b> \\
                &\nOpt{exp_1} \nChoose{a}{exp_1} \nChoose{b}{exp_1} \nPdF{z}{t^{1,s}}[S[\nBeta]]<a~b> \\
                &\nOpt{ter} \nChoose{a}{ter} \nChoose{b}{ter} \nWait{a} \nWait{b} \nClose{z}
            \nEndCase
        \end{split} \\
        \begin{split}
N^{s,s} = \nCase{z}
                &\nOpt{ins_0} \nChoose{b}{ter} \nWait{b}
                    \nRecv{r_0}{z} \nRecv{r_1}{z} \nRecv{y}{z} \\&\qquad\nCase{z}
                        \nOpt{i} \nPdF{z}{p^{s,s}_i}[\nAlpha,\nBeta]<r_0~r_1~a~y>
                    \nEndCase_{i\in L\mM} \\
                &\nOpt{ins_1} \nChoose{a}{ter} \nWait{a}
                    \nRecv{r_0}{z} \nRecv{r_1}{z} \nRecv{y}{z} \\&\qquad\nCase{z}
                        \nOpt{i} \nPdF{z}{p^{s,s}_i}[\nAlpha,\nBeta]<r_0~r_1~b~y>
                    \nEndCase_{i\in L\mM} \\
                &\nOpt{exp_0} \nChoose{a}{exp_0} \nChoose{b}{exp_0} \nPdF{z}{t^{s,s}}[S[\nAlpha],\nBeta]<a~b> \\
                &\nOpt{exp_1} \nChoose{a}{exp_1} \nChoose{b}{exp_1} \nPdF{z}{t^{s,s}}[\nAlpha,S[\nBeta]]<a~b> \\
                &\nOpt{ter} \nChoose{a}{ter} \nChoose{b}{ter} \nWait{a} \nWait{b} \nClose{z}
            \nEndCase
        \end{split}
    \end{align*}
}{}

\end{document}